\documentclass[11pt]{article}
\usepackage{hyperref}
\usepackage{fullpage}

\usepackage{color}
\usepackage{amsmath, amsfonts, amsthm, amssymb, graphicx}
\usepackage{xspace}
\usepackage{epsfig}
\usepackage{psfrag}

\newcommand{\op}{\ensuremath{\mathcal{O}}\xspace}
\newcommand{\del}{\partial} 

\newcommand{\fdel}[2][]{\ensuremath{\frac{\delta #1}{\delta #2}}}

\newcommand{\vev}[1]{\ensuremath{\langle #1 \rangle}\xspace}

\def\ie{{\it i.e.\ }}

\newcommand{\Tr}{\text{Tr}}
\newcommand{\hf}{\frac{1}{2}}
\newcommand{\qt}{\frac{1}{4}}

\let\a=\alpha \let\b=\beta \let\g=\gamma \let\d=\delta 
    
\let\l=\lambda \let\m=\mu \let\n=\nu  
\let\s=\sigma     
  \let\D=\Delta  
    
    \let\G=\Gamma

\newcommand{\be}{\begin{equation}}
\newcommand{\ee}{\end{equation}}
\def\ba{\begin{array}}
\def\ea{\end{array}}
\newcommand{\bea}{\begin{eqnarray}}
\newcommand{\eea}{\end{eqnarray}}

\newcommand{\as}[1]{_{\{ #1 \}}}
\newcommand{\asct}[1]{_{ #1 ,\text{ct}}}

\newcommand{\bs}[1]{_{( #1 )}}
\newcommand{\ab}[2]{_{\{ #1 \} (#2) }}
\newcommand{\Dp}{\Delta_+}
\newcommand{\Dm}{\Delta_-}
\newcommand{\Dpm}{\Delta_\pm}

\begin{document}

\title{Irrelevant deformations and the holographic\\ Callan-Symanzik equation}
\author{Balt C. van Rees\thanks{vanrees@insti.physics.sunysb.edu}}
\date{}
\maketitle

\begin{center}
\it C. N. Yang Institute for Theoretical Physics\\
State University of New York, Stony Brook, NY 11794-3840
\end{center}

\vskip 1cm
\begin{abstract}
We discuss the systematics of obtaining the Callan-Symanzik equation within the framework of the gauge/gravity dualities. We present a completely general formula which in particular takes into account the new holographic renormalization results of \cite{me}. Non-trivial beta functions are obtained from new logarithmic terms in the radial expansion of the fields. The appearance of multi-trace counterterms is also discussed in detail and we show that mixing between single- and multi-trace operators leads to very specific non-linearities in the Callan-Symanzik equation. Additionally, we compute the conformal anomaly for a scalar three-point function in a CFT.
\end{abstract}

\newpage
\tableofcontents
\newpage

\section{Introduction}
In the past decade the gauge/gravity dictionary has been developed with ever-increasing precision. At the supergravity level, one of its essential ingredients is the framework of \emph{holographic renormalization} (see \cite{Skenderis:2002wp} for a review) which is the process of regularizing and renormalizing the infrared divergences in the on-shell gravity action. This paves the way for the holographic computation of renormalized field theory correlation functions and furthermore leads directly to specific functional constraints for the renormalized on-shell action which precisely correspond to the Ward identities of the dual field theory.
 
In the implementation of holographic renormalization it is generally assumed that the bulk spacetime is of a so-called \emph{Asymptotically locally AdS} (or AlAdS) form. This implies that near the conformal boundary $r \to \infty$ the metric can be put in the Fefferman-Graham form \cite{FeffermanGraham,Graham:1999jg}:
\be
ds^2 = dr^2 + \g_{ij} dx^i dx^j \qquad \qquad \g_{ij} = e^{2r} g_{(0)ij} + \ldots
\ee
where $g_{(0)ij}$ is a non-degenerate boundary metric and the dots represent subleading terms as $r \to \infty$. This property of the spacetime indicates that the dual field theory is UV conformal and thus guarantees its renormalizability by standard quantum field theory reasoning. Correspondingly the on-shell bulk action for AlAdS spacetimes should be holographically renormalizable as well and this was indeed demonstrated quite generally in \cite{Papadimitriou:2005ii}.

\paragraph{Extension to non-AlAdS spacetimes\\}
In recent years a variety of conjectures has been made concerning a possible holographic interpretation of spacetimes which are \emph{not} of an AlAdS form, see for example \cite{Hashimoto:1999ut,Maldacena:1999mh,Son:2008ye,Balasubramanian:2008dm,Kachru:2008yh,Kanitscheider:2008kd}. In general the holographic dictionary is far less well-developed for such spacetimes, in particular the extension of the holographic renormalization procedure has proved to be all but straightforward. Unfortunately, this makes it difficult to obtain general holographic results like Ward identities or perform precision tests of these dualities. (A notable exception are the non-conformal branes of \cite{Kanitscheider:2008kd,Kanitscheider:2009as}.) 

It is actually quite natural to expect difficulties in extending the holographic renormalization procedure to non-AlAdS spacetimes. The modified asymptotic structure of the spacetime indicates that the dual theory is no longer UV conformal and therefore renormalizability should indeed become a genuine issue both in the bulk and on the boundary. In fact, in many cases the modified bulk asymptotics can be directly interpreted as being due to specific \emph{irrelevant} deformations of the dual field theory \cite{Son:2008ye,us,Costa:2010cn} which makes the issue of renormalizability very explicit. Notice that some encouraging results were nevertheless obtained in \cite{us,Costa:2010cn} where it was found that the non-renormalizability appeared to be rather restricted and therefore the holographic analysis still gave physically meaningful results. 

In the recent paper \cite{me} we initiated a systematic study of holography for non-AlAdS spacetimes by investigating \emph{perturbatively small} irrelevant deformations of standard AlAdS backgrounds. By working only to finite order in the sources the field theory remained fully renormalizable and correspondingly we found that the on-shell bulk action could be completely holographically renormalized as well. Within this setting we were able to uncover several new and rather nontrivial structures which we expect to be directly relevant for the holographic interpretation of non-AlAdS spacetimes.

The main results of \cite{me} are the following. First of all, we found new terms in the radial expansion of the bulk fields which are \emph{more} leading than the terms corresponding to the field theory sources. The presence of these terms implies that the sources become non-linear functions of the asymptotics of the bulk fields. We also found so-called \emph{pseudo-non-local} divergences, which are divergences that are non-local in the sources and should have appeared by naive power-counting, but happen to vanish once we include merely local counterterms. Finally we wrote down counterterms involving the conjugate momentum of the bulk fields which are of a qualitatively new type and correspond to \emph{multi-trace} counterterms in the dual field theory. We emphasize that these results are applicable to all holographic computations of correlation functions involving irrelevant operators and thus form an integral part of the usual AdS/CFT dictionary.

\paragraph{Logarithmic divergences\\}
In \cite{me} we for simplicity picked models with scaling dimensions $\D$ such that only power-law divergences appeared, \ie the divergences were never logarithmic in the cutoff. This in particular implied that the renormalization procedure did not introduce any anomalous behavior under scale transformations and therefore the naive Callan-Symanzik equation remained unaffected. However in realistic gauge/gravity dualities one does frequently encounter such logarithmic divergences. The aim of this paper is therefore to repeat the analysis of \cite{me} precisely for these logarithmic cases and to find out how the computation of the holographic Callan-Symanzik equation is affected by the new results of \cite{me} which we summarized in the previous paragraph.

Below we study this problem in the following way. First we review some standard aspects of renormalization in quantum field theory in section \ref{sec:ren}. We begin by presenting the Callan-Symanzik equation in its most general form and subsequently discuss some of its properties when the field theory is conformally invariant. In anticipation of the gravity results we also study mixing between single- and multi-trace operators at large $N$.

Afterwards, in section \ref{sec:logs}, which is the main section of the paper, we study in detail the same toy $\lambda \Phi^3$ model as in section 2 of \cite{me}. This time however we will pick a scaling dimension $\D$ such that in particular the multi-trace divergence becomes of a logarithmic form and we will then focus on the effects of the counterterms on the holographic Callan-Symanzik equation. We find that the new structures mentioned above indeed lead to various unconventional features which we will describe in detail below. After disentangling all the issues we eventually obtain a Callan-Symanzik equation which precisely matches the field theory results of section \ref{sec:ren}.

From the example in section \ref{sec:logs} we may distill a general method for obtaining the holographic Callan-Symanzik equation. This method is presented in section \ref{sec:concl} and we consider it the main result of this paper. In this final section we also discuss the relation of our work with the holographic renormalization literature and mention some applications and open problems.

Finally, a byproduct of our analysis is the holographic computation of the conformal anomaly for a scalar three-point correlation function which we present in \eqref{eq:preciseanomaly3pt}. As far as we are aware this is a new result whose universality is guaranteed by the non-renormalization of \cite{Petkou:1999fv}. It would be interesting to obtain a Weyl-covariant version as in equation (17) of \cite{Papadimitriou:2004ap}.

\section{Aspects of renormalization in conformal field theories}
\label{sec:ren}
In this section we review some general aspects of renormalization in quantum field theory. We begin with a discussion of the Callan-Symanzik equation at the level of the partition function. We then focus on conformal field theories and give a detailed treatment of the renormalization of scalar two- and three-point functions, leading to the appearance of conformal anomalies for certain scaling dimensions. Finally we consider operator mixing between single- and multi-trace operators and discuss how this leads to specific non-linearities in the Callan-Symanzik equation at large $N$. 

\subsection{The general Callan-Symanzik equation}
In this paper we write the Callan-Symanzik equation as a functional differential equation for the renormalized partition function $W[t^K]$, which is by definition the generator of connected correlation functions,
\be
\label{eq:defnW}
\exp(-W[t^K]) = \int D\Phi \exp\Big(-S[\Phi] - \int d^d x \,\sum_K t^K(x) \op_K(x) \Big)\,.
\ee
The partition function depends on the various sources labelled $t^K$ for operators $\op_K$, where $K$ is an abstract index labelling the different operators in the theory and may include Lorentz indices. As is explained for example in \cite{Osborn:1991gm}, general quantum field theory expectations dictate that the Callan-Symanzik equation takes the form:
\be
\label{eq:generalcseqn}
\begin{split}
\m \frac{\del}{\del \m} W[t^K] + \int d^d x \, \sum_I \b^I[t^K(x)] \frac{\d W[t]}{\d t^I(x)} + \int d^d x \mathcal A[t(x)] = 0\,,
\end{split}
\ee
where $\mu$ is the renormalization scale, $\mathcal A$ is the conformal anomaly of the theory and the $\b^I$ reduce to the standard beta functions of the theory if we restrict ourselves to constant sources. For an ordinary renormalizable quantum field theory, $\b^I$ and $\mathcal A$ are local functions of the sources $t^K(x)$, at least when expanded to any finite order in the number of sources. Upon functional differentiation with respect to the sources one obtains the Callan-Symanzik equation in its familiar form \cite{Callan:1970yg,Symanzik:1970rt}, namely as a differential equation for the renormalization scale dependence of connected correlation functions.

Notice that $W[t^K]$ is dimensionless and therefore its $\m$-derivative should be equal to a rescaling of all distances combined with a rescaling of all the dimensionful sources. We may implement the rescaling of all distances by a rescaling of the metric $g_{ij}$ on which $W[t^K]$ also depends. We then obtain that \cite{Osborn:1991gm,mythesis}
\be
\label{eq:mddmW}
\m \frac{\del}{\del \m} W[t^K] = \Big(\int d^d x \, 2 g_{ij} \fdel{g_{ij}} + \int d^d x \, \sum_I (\D_I - d - \tilde s_I) t^I(x) \fdel{t^I(x)}\Big) W[t^K]\,,
\ee
where $\tilde s_I$ is the number of upper minus the number of lower vector Lorentz indices on the source $t^I(x)$ and $\D_I$ is the scaling dimension of $\op_I(x)$ in the theory with all the sources $t^K(x)$ set to zero.

In section \ref{sec:logs} we will see that the holographic computation will precisely yield a Callan-Symanzik equation of the form \eqref{eq:generalcseqn} with the $\mu$-derivative replaced with \eqref{eq:mddmW}.

\subsection{Renormalization of the two-point function}
\label{sec:2ptfn}
Consider a $d$-dimensional CFT with a conformal primary scalar operator $\op$ of scaling dimension $\D$. It is well-known that conformal invariance fixes the two-point function in flat space to be of the form:
\be
\label{eq:cft2pt}
\vev{\op(x)\op(0)} = \frac{c_2}{x^{2\D}}\,.
\ee
The right-hand side as it stands is however not valid at the contact point $x = 0$. This is because its integral does not converge at $x=0$ for $\D \geq d/2$ and therefore \eqref{eq:cft2pt} is not well-defined in the distributional sense. On the other hand, in quantum field theory we do expect correlation functions to be well-defined distributions, so we can for example integrate them against sources (to obtain a finite partition function) or take their Fourier transform. Indeed, in ordinary perturbation theory the standard renormalization procedures automatically result in such finite distributions. The correct two-point function is therefore not quite given by equation \eqref{eq:cft2pt} but rather some renormalization procedure is needed to find a finite distribution which reduces to \eqref{eq:cft2pt} away from contact points. In the remainder of this subsection we will review the computation of \cite{Petkou:1999fv} (see also \cite{Osborn:1993cr}) where the principles of differential regularization \cite{Freedman:1991tk} were used to obtain such a correctly renormalized two-point function.

Let us first consider the case where $\D - d/2 \notin \{0,1,2,\ldots\}$. In those cases we can recursively use
\be
\begin{split}
\frac{1}{x^{2\D}} = \frac{1}{4(\D - d/2) (\D -1)} \square \frac{1}{x^{2\D -2}}
\end{split} 
\ee
to rewrite
\be
\label{eq:x2deltaren}
\frac{1}{x^{2\D}} =
\frac{\G(\D - k + 1 - d/2) \G(\D -k)}{4^k \G(\D - d/2 + 1)\G(\D)} \square^k \frac{1}{x^{2\D - 2k}}\,, 
\ee
with $k$ the smallest integer such that $k > \D - d/2$. We then define a finite `renormalized' correlation function, denoted
\be
\mathcal R \frac{1}{x^{2\D}},
\ee
as the distribution obtained via a formal integration by parts:
\be
\label{eq:ibpdistr}
\int d^d x f(x) \mathcal R \frac{1}{x^{2\D}} \equiv \frac{\G(\D - k + 1 - d/2) \G(\D -k)}{4^k \G(\D - d/2 + 1)\G(\D)} \int d^d x \frac{1}{x^{2\D - 2k}} \square^k f(x)\,.  
\ee
Indeed, the right-hand side of \eqref{eq:ibpdistr} is finite at $x = 0$ for any sufficiently regular test function $f(x)$. With a slight abuse of notation we will henceforth write:
\be
\label{eq:rencorr}
\mathcal R \frac{1}{x^{2\D}} = \frac{\G(\D - k + 1 - d/2) \G(\D -k)}{4^k \G(\D - d/2 + 1)\G(\D)} \square^k \frac{1}{x^{2\D - 2k}}\,, \qquad \qquad \D - d/2 \notin \mathbb \{0,1,2,\ldots\}\,,
\ee
and a procedure like \eqref{eq:ibpdistr} is then implicitly understood.

In the special case where $\D - d/2 = n$ with $n \in \{0,1,2,\ldots\}$ the above prescription does not quite suffice and logarithmic divergences remain. For those cases we replace, for $d \neq 2$,
\be
\frac{1}{x^d} \to \frac{-1}{2(d-2)} \square \frac{\ln(x^2 \m^2)}{x^{d-2}}\,,
\ee
so that \eqref{eq:x2deltaren} becomes
\be
\label{eq:rencorr2}
\mathcal R \frac{1}{x^{2n + d}} = \frac{-\G(d/2)}{2^{2n+1} (d-2)\G(n+1)\G(n+d/2)} \square^{n+1} \frac{\ln(x^2 \m^2)}{x^{d-2}}\,, \qquad \qquad d \neq 2\,.
\ee
Finally, for $d=2$ and still $\D - d/2 = n$ one uses
\be
\frac{1}{x^{2}} \to \frac{1}{8} \square \ln^2(x^2 \m^2) 
\ee
to obtain
\be
\label{eq:rencorr3}
\mathcal R \frac{1}{x^{2n+2}} = \frac{1}{2^{2n+3} \G(n+1)^2} \square^{n+1} \ln^2(x^2 \m^2)\,, \qquad \qquad d=2\,.
\ee

In summary, the distributions defined in \eqref{eq:rencorr}, \eqref{eq:rencorr2} and \eqref{eq:rencorr3} allow us to write the correctly renormalized two-point function for any $\D$ and $d$ as:
\be
\vev{\op(x)\op(0)} = \mathcal R \frac{c_2}{x^{2\D}}\,.
\ee
As promised, such correlation functions are functionally equal to $x^{-2\D}$ away from contact points but they are also well-defined distributions at $x = 0$.

\paragraph{Conformal anomalies\\}
Although the above discussion may appear to be somewhat pedantic, for the special values $\D - d/2 = n$ it has definite physical consequences in the form of a nontrivial dependence on a renormalization scale $\mu$. Indeed, we now find that in those cases (for any $d$):
\be
\m \frac{d}{d \m} \vev{\op(x)\op(0)} =
\frac{c_2 \pi^{d/2}}{2^{2n-1} \G(n+1)\G(n+d/2)} \square^{n}\d^d(x)\,,
\ee
where we used that
\be
\square \frac{1}{x^{d-2}} = \frac{2(2-d)\pi^{d-2}}{\G(d/2)}\d^d(x)
\ee
as well as
\be
\square \ln(x^2 \m^2) = 4 \pi \d^2(x)
\ee
which is only valid for $d = 2$.

\paragraph{Scheme dependence\\}
The renormalized correlation functions we defined above are generally not unique. We may for example add delta functions or derivatives of delta functions to the correlation function to obtain a different but still finite result with the right conformal dimension. For example, for $\D = n + d/2$ we can change the renormalization scale $\m$ to obtain a new renormalized correlation function of the form:
\be
\mathcal R \frac{1}{x^{2n + d}} +  a_2 \square^n \d^d(x)\,,
\ee
where $a_2$ can be an arbitrary finite dimensionless coefficient. This scheme dependence does not modify the conformal anomaly of the two-point function. However the conformal anomaly of the three-point function, which we compute below, is partially dependent on the scheme chosen for the two-point function. 

\subsection{Renormalization of the three-point function}
Our next example is the renormalization of the three-point function of three identical scalar operators. Away from contact points it takes the form:
\be
\vev{\op(x)\op(y)\op(z)} = \frac{c_3}{(x-y)^\D (y-z)^\D (z-x)^\D}\,.
\ee
Divergences arise when either two out of the three points coincide or when all three points coincide. Rather than regulating all these divergences let us for the moment focus on the case where $x \to y$ only and compute the associated conformal anomaly. The relevant part of the operator product expansion is given by:
\begin{multline}
\op(x)\op(0) \sim \frac{c_2}{x^{2\D}} + \frac{c_3}{c_2 x^\D} \Big[ 
1 + \hf x^\m \del_\m + \frac{\D + 2}{8(\D + 1)} x^\m x^\n \del_\m \del_\n\\  - \frac{\D}{8(\D + 1)(2\D - d + 2)}x^2 \square + \ldots \Big] \op(0) + \ldots
\label{eq:ope}
\end{multline}
where we included the contribution of the first two descendants. We now substitute the OPE in the three-point function, use the fact that $\vev{\op(x)}$ vanishes by conformal invariance and replace the singular distributions as $x \to 0$ by renormalized distributions to find that:
\begin{multline}
\label{eq:threeptren}
\vev{\op(x)\op(0)\op(z)} = \frac{c_3}{c_2} \Big[ \Big(\mathcal R \frac{1}{x^\D}\Big)
- \hf \Big( \mathcal R \frac{x^\m}{x^\D}\Big) \del_\m + \frac{\D + 2}{8(\D + 1)} \Big(\mathcal R \frac{x^\m x^\n}{x^\D}\Big) \del_\m \del_\n\\  - \frac{\D}{8(\D + 1)(2\D - d + 2)} \Big(\mathcal R \frac{1}{x^{\D - 2}}\Big) \square + \ldots \Big] \vev{\op(z)\op(0)} + \ldots
\end{multline}
where the distribution corresponding to $x^\m/x^\D$ is defined for $\D \neq 2$ as:
\be
\mathcal R \frac{x^\m}{x^\D} =
\frac{-1}{\D -2} \del^\m \mathcal R \frac{1}{x^{\D - 2}}
\ee
and the derivative of the renormalized distribution on the right-hand side is as usual defined via a formal integration by parts. For $\D = 2$ we may use
\be
\mathcal R \frac{x^\m}{x^2} = \frac{1}{2}\del^\m \log(x^2 \m^2)\,.
\ee
We also define
\be
\mathcal R \frac{x^\n x^\m}{x^{\D}} = \frac{1}{\D - 2} \Big(\frac{1}{\D - 4} \del^\m \del^\n \mathcal R \frac{1}{x^{\D - 4}} +  \delta^{\m \n} \mathcal R \frac{1}{x^{\D - 2}} \Big)
\ee
for $\D \neq 2,4$. For $\D = 2$ no renormalization is needed and for $\D = 4$ we use
\be
\mathcal R \frac{x^\m x^\n}{x^4} = - \frac{1}{4}\del^\m \del^\n \ln(x^2 \m^2) + \frac{1}{2} \d^{\m \n} \mathcal R \frac{1}{x^2}\,.
\ee
We have now defined all the renormalized distributions appearing in \eqref{eq:threeptren}. From the above results one may directly compute the associated conformal anomaly for any $\D$ or $d$. We will not give the general result here but instead restrict ourselves to two specific examples.

First let us consider the case where $\D =d$. In that case we find that
\be
\label{eq:marginalanomaly}
\m \frac{d}{d\m}\vev{\op(x)\op(y)\op(z)} = \frac{ 2 c_3 \pi^{d/2}}{c_2 \G(d/2)}\d^d(x-y) \vev{\op(y) \op(z)} + (\text{permutations}) + (\text{ultralocal})\,,
\ee
where the terms we omitted are either permutations of $(x,y,z)$ of the first term or ultralocal terms which are nonzero only when all three points come together. 

Our second example is the case where $d=2$ and $\D =4$ which we will match to a holographic computation below. In that case we find that
\begin{multline}
\label{eq:threeptanomaly}
\m \frac{d}{d\m} \vev{\op(x)\op(0)\op(z)} = \frac{\pi c_3}{2 c_2}\Big[
(\square \d^2(x)) + (\del^\m \d^2(x)) \del_\m + \qt \d^2(x) \square 
 \Big] \vev{\op(z)\op(0)} \\
 + (\text{permutations}) + (\text{ultralocal})\,.
\end{multline}
As far as we are aware this conformal anomaly for three scalar operators has not been computed before in the literature and it would be interesting to extend the above methods to also obtain the ultralocal component of the anomaly. We will instead compute it using holographic methods in the next section.

Notice that in both of the above examples all the subsequent terms that appear in the operator product expansion \eqref{eq:ope} are finite and therefore do not give rise to further conformal anomalies.

\subsection{Conformal anomalies and the partition function}
\label{sec:confanomaliespartfn}
Let us now integrate the correlation functions against sources to obtain the anomalous transformation properties of the partition function $W[t]$ which we defined in \eqref{eq:defnW}. We henceforth consider only a single operator $\op$ and therefore we will drop the index $K$ which appeared in \eqref{eq:defnW}. By expanding in the source $t(x)$ we find that up to the level of the three-point function the partition function is given by:
\be
\begin{split}
W[t] &= - \hf \int d^d x d^d y \, \vev{\op(x)\op(y)} \, t(x) t(y) \\ &\qquad + \frac{1}{6} \int d^d x d^d y d^d z \, \vev{\op(x)\op(y)\op(z)}\, t(x) t(y) t(z)  + \ldots
\end{split}
\ee

Let us focus on the case where $d=2$ and $\D = 4$. Using the results of the previous section we straightforwardly find that
\be
\begin{split}
\m \frac{\del}{\del \m} W[t] &= \frac{\pi c_3}{4 c_2}  \int d^2 x d^2 z  \Big[
t(x) \square t(x) - \qt \square t(x)^2
 \Big] \vev{\op(x)\op(z)} t(z)\\
 & \qquad - \int d^2 x \Big[ \frac{c_2 \pi}{64 \cdot 36} t(x) \square^3 t(x) + (\text{ultralocal}) \Big] + \ldots
 \end{split}  
\ee
where we again did not explicitly write the ultralocal terms in the anomalous behavior of the three-point function. We may alternatively write this as follows:
\be
\label{eq:CSmddmW}
\begin{split}
\m \frac{\del}{\del \m} W[t] &= - \frac{\pi c_3}{4 c_2}  \int d^2 x  \Big[
t(x) \square t(x) - \qt \square t(x)^2 
 \Big] \frac{\d W[t]}{\d t(x)}\\
 & \qquad - \int d^2 x \Big[ \frac{c_2 \pi}{64 \cdot 36} t(x) \square^3 t(x) + (\text{ultralocal}) \Big] + \ldots
 \end{split}
\ee
This equation is precisely of the form \eqref{eq:generalcseqn} and we can directly read off that
\be
\beta(t) = \frac{\pi c_3}{4 c_2} \Big[ t(x) \square t(x) - \qt \square t(x)^2 \Big] \,.
\ee
Notice that this beta function vanishes for constant $t(x)$. This would not be the case for a marginal operator with $\D = d$ and $c_3 \neq 0$ as can be seen directly from \eqref{eq:marginalanomaly}. 

In the second part of this paper we will compare equation \eqref{eq:CSmddmW} with the results from the holographic renormalization procedure and find perfect agreement. Furthermore, we will also obtain the ultralocal piece at the level of the three-point function which we present in \eqref{eq:preciseanomaly3pt} below.

\subsection{Multi-trace counterterms and operator mixing}
\label{sec:multitrace}
In this subsection we consider mixing between single- and multi-trace operators in the large $N$ limit. As we review in the appendix to this paper, at large $N$ the multi-trace partition function
\be
\exp(- N^2 w[t, f]) = \int D\Phi \exp\Big(- S - N \int t\, \op - N^2 \int f(N^{-1} \op) \Big)
\ee
takes the form
\be
\label{eq:wmultitracemain}
w[t,f] = w_0[t + f'(\s)] + \int d^d x (f(\s) - f'(\s) \s)\,,
\ee
where
\be
\label{eq:smultitracemain}
\s = w'_0[t + f'(\s)]
\ee
and $w_0[t] = w[t,0]$ is simply the single-trace partition function. We take the multi-trace deformation $f(\s)$ to be of the form:
\be
f(\s(x)) = \sum_{k \geq 2} t_k(x) \s(x)^k
\ee
and we also define $t_1(x) = t(x)$ so the functions $t_k(x)$ source the operator $\op^k$ for all $k \geq 1$. Upon a small variation of the sources $t_k(x)$ we obtain that
\be
\label{eq:deltawmultitracemain}
\delta w[t_k] = \sum_{k \geq 1}\int d^d x \, \s(x)^k \delta t_k(x)
\ee
and the one-point functions of $\op^k$ are therefore given as:
\be
\vev{\op^k(x)} = N^k \sigma(x)^k\,.
\ee
Let us now consider the Callan-Symanzik equation for multi-trace operators at large $N$. If the only nonzero sources are the $t_k(x)$ then the general Callan-Symanzik equation \eqref{eq:generalcseqn} takes the form:
\be
\m \frac{\del}{\del \m} w[t_k] + \int d^d x \, \sum_{i \geq 1}\b_i[t_k (x)] \frac{\d w[t]}{\d t_i(x)} + \int d^d x \mathcal A[t_k(x)] = 0\,,
\ee
where we absorbed an extra $N^{-2}$ in the definition of the conformal anomaly. Upon substitution of \eqref{eq:wmultitracemain} and using \eqref{eq:smultitracemain} and \eqref{eq:deltawmultitracemain} we find that we may rewrite this equation as:
\be
\label{eq:mddmwmultitr}
\m \frac{\del}{\del \m} w_0\Big[\sum_{k \geq 1}k t_k \s^{k-1}\Big] + \int d^d x \, \sum_{i \geq 1} \b_i[t_k(x)] \s(x)^i + \int d^d x \mathcal A[t_k(x)]=0\,.
\ee 
Notice that in particular all the terms involving $\m \del_\m \s$ cancel which is why we recover merely the $\m$-variation of the single-trace partition function. As in \eqref{eq:wmultitracemain} it should be evaluated at the shifted source
\be
t + f'(\s) = \sum_{k \geq 1}k t_k \s^{k-1}\,.
\ee 
We will now work out equation \eqref{eq:mddmwmultitr} a bit further in two examples.

\paragraph{Marginal double-trace operator\\}
Our first example is a scalar operator with $\D = d/2$. In that case we may use the results of section \ref{sec:ren} to find that the conformal anomaly at order $t(x)^2$ for the single-trace partition function is given by: 
\be
\m \frac{\del}{\del \m} w_0[t] = - \int d^d x \mathcal A[t] =  - \int d^d x \frac{2 c_2 \pi^{d/2}}{\G(d/2)} t(x)^2\,.
\ee 
For a shifted source we therefore directly obtain that:
\be
\m \frac{\del}{\del \m} w_0[t_1 + 2 t_2 \s ] + \int d^d x \frac{2 c_2 \pi^{d/2}}{\G(d/2)} (t_1(x) + 2 t_2(x) \s(x))^2 = 0\,.
\ee
Comparison with \eqref{eq:mddmwmultitr} yields once more the expected expression for the conformal anomaly $\mathcal A$ but we now also find the non-trivial beta functions: 
\be
\begin{split}
\b_1 &= \frac{8 c_2 \pi^{d/2}}{\G(d/2)} t_1 t_2\\
\b_2 &= \frac{8 c_2 \pi^{d/2}}{\G(d/2)} t_2^2\,,
\end{split}
\ee
for the single- and double-trace operator, respectively. We may also switch off the single-trace interaction and obtain a Callan-Symanzik equation just for the double-trace deformation:
\be
\label{eq:betadoubletrace}
\Big(\m \frac{\del}{\del \m}  + \int d^d x \frac{8 c_2 \pi^{d/2}}{\G(d/2)} t_2^2  \fdel{t_2}\Big) w[t_2] = 0\,.
\ee
We conclude that the conformal anomaly in the two-point function of an operator $\op$ of dimension $d/2$ directly results in the above beta function for $\op^2$. Furthermore, if we restrict ourselves to only a double-trace deformation then it is not hard to deduce that there cannot be any higher-order contributions to \eqref{eq:betadoubletrace}. The beta function obtained from \eqref{eq:betadoubletrace} is therefore exact in $t_2$, a result which is in agreement with the diagrammatic analysis of \cite{Pomoni:2008de}.

\paragraph{Irrelevant operators\\}
Our second example considers operator mixing between irrelevant single- and multi-trace operators. Let us suppose that the single-trace operator $\op$ mixes with the double-trace operator $\op^2$ at order $k$ in the sources. As is explained for example in the textbook \cite{zinnjustin}, at the level of correlation functions this means that
\be
\op(x_1) \op(x_2) \op(x_3) \ldots \op(x_k)
\ee
mixes with
\be
\delta^d (x_1 - x_2) \delta^d(x_2 - x_3) \ldots \delta^d(x_{k-1} - x_{k}) \op^2(x_k)\,.
\ee
At the level of the partition function this implies that we find a nonzero $\beta$ function for $\op^2$ of the form:
\be
\b_2 = \alpha t_1(x)^k\,,
\ee 
with $\alpha$ a constant. Notice that by power-counting such a term can only arise if $(k-2) \D = (k-1) d$, so in particular we need $\D > d$ and $k \geq 3$. Suppose also that we a priori do not switch on any sources for the double-trace operator and let us ignore all other beta functions and conformal anomalies. In that case equation \eqref{eq:mddmwmultitr} combined with \eqref{eq:mddmW} reduces to the following equation for the single-trace partition function:
\begin{multline}
\label{eq:nonlincs}
\Big(\int d^d x \, 2 g_{ij} \fdel{g_{ij}}  + \int d^d x \frac{d}{k-2} t(x)  \fdel{t(x)}\Big)  w_0[t] \\ = \a \int d^d x \, t(x)^k \s(x)^2 = \a \int d^d x \, t(x)^k \Big(\frac{\d w_0[t]}{\d t(x)}\Big)^2\,,
\end{multline}
where for later reference we re-wrote $\sigma(x)$ as the variation of the partition function $w_0[t]$. This perhaps un-intuitive Callan-Symanzik equation, which is non-linear in first derivatives, is precisely the equation we will obtain holographically below. However we may now conclude that its proper field theory interpretation is simply given in terms of a standard Callan-Symanzik equation involving single- and double-trace operator mixing at large $N$.

We note that multi-trace counterterms generically make their appearance if one does not send the field theory cutoff to infinity. Correspondingly, one may expect to encounter equations similar to \eqref{eq:nonlincs} in the planar version of Polchinski's renormalization group equation, or alternatively in the study of holography with a finite cutoff. This is indeed the case and was studied in \cite{Becchi:2002kj,Li:2000ec,Heemskerk:2010hk,Akhmedov:2002gq}. In this paper we do send the cutoff to infinity and consider the appearance of multi-trace operator mixing for the fully renormalized partition function.

\section{Scalar field with single-trace counterterms}
\label{sec:logs}
In this section we work out the holographic renormalization of the on-shell action for a massive scalar field $\Phi$ with a $\lambda \Phi^3$ interaction, order by order in $\lambda$. This is the same toy model as in section 2 of \cite{me}, but this time we choose the specific values $d=2$ and $\D = 4$, so $m^2 = \D (\D - d) = 8$. Choosing these values makes it precisely one of the cases we excluded in \cite{me} because the radial expansion of $\Phi$ is no longer a simple power series in $\exp(-r)$ but rather involves `logarithmic' terms which are well-known to complicate the usual holographic renormalization analysis. (In our coordinate system these `logarithmic' terms are actually of the form $\log(e^{-r}) = -r$, so polynomial in $r$.) Notice that we specifically chose $\D = 2d$ in order for the multi-trace divergence exhibited in \cite{me} to be of logarithmic type.

The computations in this section are largely analogous to those presented in section 2 of \cite{me}. We will therefore assume some familiarity with the contents of that paper and focus on the differences brought about by the logarithmic terms.

\subsection{Setup}
In this subsection we set up the bulk problem, following \cite{me}. The three-dimensional bulk action is given by:
\be
S = \int d^3 x \sqrt{G} \Big( \hf \del_\m \Phi \del^\m \Phi + 4 \Phi^2 + \frac{1}{3} \lambda \Phi^3\Big)\,.
\ee
The background spacetime is empty Euclidean AdS$_3$ with a metric of the form:
\be
\label{eq:bgmetric}
G_{\m \n}dx^\m dx^\n = dr^2 + \g_{ij} dx^i dx^j \qquad \qquad \g_{ij} = e^{2r} \d_{ij}\,,
\ee
with $i \in \{1, 2\}$. The equation of motion becomes:
\be
\label{eq:eomphi}
\square_G \Phi - 8 \Phi - \lambda \Phi^2 = \ddot \Phi + 2 \dot \Phi + e^{-2r} \square_0 \Phi - 8 \Phi - \lambda \Phi^2 = 0\,,
\ee
where a dot denotes a radial derivative and we introduced $\square_0 = \delta^{ij} \del_i \del_j$. We will below also use the covariant Laplacian $\square_\g = \g^{ij} \del_i \del_j = e^{-2r} \d^{ij} \del_i \del_j$. We will solve equation \eqref{eq:eomphi} only perturbatively in $\lambda$, ignoring the backreaction onto the metric or any other fields as well as any non-perturbative effects. To this end we expand the solution $\Phi$ as:
\be
\label{eq:phifullexp}
\Phi = \Phi\as{0} + \Phi\as{1} + \Phi\as{2} + \Phi\as{3} + \ldots
\ee   
with the individual terms given by the solutions to linear equations:  
\be
\label{eq:eomseries}
\begin{split}
&(\square_G - 8) \Phi\as{0} = 0\\
&(\square_G - 8) \Phi\as{1} = \lambda \Phi^2\as{0}\\
&(\square_G - 8) \Phi\as{2} = 2 \lambda \Phi\as{0} \Phi\as{1}\\
&(\square_G - 8) \Phi\as{3} = \lambda ( \Phi\as{1}^2 + 2 \Phi\as{0} \Phi\as{2})\\
&\ldots
\end{split}
\ee
and boundary conditions which say that
\be
\Phi = \ldots + e^{2r} \phi\ab{0}{-2} + \ldots
\ee
These boundary conditions are explained in more detail in \cite{me}. They will be unaffected by the presence of logarithmic terms.

The on-shell action takes the form:
\be
\label{eq:sonshell}
S = - \frac{\lambda}{6} \int d^3 x \sqrt{G} \Phi^3 + \hf \int d^2 x \sqrt{\g} \dot \Phi \Phi \,,
\ee
where for our choice of background coordinate system we find that $\sqrt{G} = \sqrt{\g} = e^{2 r}$.

\subsection{Free-field analysis and two-point function}
\label{sec:freefield}
The first step is to perform the holographic renormalization at order $\lambda^0$. At this level the scalar field is free and the analysis is completely familiar. It was first presented in \cite{deHaro:2000xn} and is also reviewed in the lecture notes \cite{Skenderis:2002wp}. We summarize the essential features below.

\paragraph{Solution of the equation of motion\\}
We begin by asymptotically solving the first equation in \eqref{eq:eomseries}, which more explicitly reads:
\be
\label{eq:freeeom}
\ddot \Phi\as{0} + 2 \dot \Phi\as{0} + e^{-2r} \square_0 \Phi\as{0} - 8 \Phi\as{0} = 0\,.
\ee

At the risk of being somewhat pedantic we will now explicitly demonstrate how this equation leads to logarithmic terms in the radial expansion of $\Phi$. We begin with the leading behavior of $\Phi$ as $r \to \infty$. Upon substitution of the ansatz $\Phi\as{0} \sim \phi\ab{0}{\a} (x^i)\exp(- \a r)$ in \eqref{eq:freeeom} we find that the allowed solutions have $\a = -2$ and $\a = 4$. Near the boundary we therefore find the two asymptotically independent solutions:
\be
\Phi\as{0} = e^{2r} \phi\ab{0}{-2} + \ldots + e^{-4r} \phi\ab{0}{4} + \ldots
\ee
The subleading terms in the expansion appear because of the explicit $\exp(-2r)$ multiplying the box in \eqref{eq:freeeom}, suggesting a radial expansion of the form:
\be
\label{eq:phirad0naive}
\Phi\as{0} = e^{2r} \phi\ab{0}{-2} + \phi\ab{0}{0} + e^{-2 r} \phi\ab{0}{2} + e^{-4r} \phi\ab{0}{4} + e^{-6r} \phi\ab{0}{6} + \ldots
\ee
Upon substitution of this ansatz into the equation of motion one finds that:
\be
\label{eq:phirad0coeff}
\begin{split}
\square \phi\ab{0}{-2} - 8 \phi\ab{0}{0} &= 0\\
\square \phi\ab{0}{0} - 8 \phi\ab{0}{2} &= 0\\
\square \phi\ab{0}{2} &=0\\
\square \phi\ab{0}{4} + 16 \phi\ab{0}{6} &= 0\,.
\end{split}
\ee
The first, second and last of these equations can be used to determine the coefficients $\phi\ab{0}{0}$, $\phi\ab{0}{2}$ and $\phi\ab{0}{6}$ in terms of the asymptotically independent boundary data $\phi\ab{0}{-2}$ and $\phi\ab{0}{4}$. However the third equation is problematic as it would lead to the constraint that $\square^3 \phi\ab{0}{-2} = 0$. Such constraints should not arise for a linear second-order differential equation and we conclude that the naive expansion \eqref{eq:phirad0naive} does not capture the most general solution to the equation of motion. The correct radial expansion is easily found to be:
\be
\label{eq:phirad0}
\Phi\as{0} = e^{2r} \phi\ab{0}{-2} + \phi\ab{0}{0} + e^{-2 r} \phi\ab{0}{2} + e^{-4r} (r \tilde \phi\ab{0}{4} + \phi\ab{0}{4}) + e^{-6r} \phi\ab{0}{6} + \ldots
\ee
Indeed, this expansion alleviates the problem and replaces the third equation in \eqref{eq:phirad0coeff} with:
\be
\tilde \phi\ab{0}{4} = \frac{1}{384} \square^3 \phi\ab{0}{-2}
\ee
and the coefficient $\phi\ab{0}{4}$ is still left undetermined by the asymptotic radial expansion.

The asymptotic solution to the equation of motion at higher orders in $\lambda$, which we present without derivation below, will feature more logarithmic terms. These either originate directly from logarithmic terms at lower orders or they are obtained in precisely the same way as above, \ie by requiring the existence of two asymptotically independent solutions.     

\paragraph{Counterterm action and one-point function\\}
The next step is to substitute the solution \eqref{eq:phirad0} into the on-shell action \eqref{eq:sonshell} at order $\lambda^0$ and find a counterterm action $S\asct{0}$ which cancels the divergences. This counterterm action can be obtained using standard methods \cite{Skenderis:2002wp} and we will not repeat its derivation here. The full counterterm action is given by:
\be
\label{eq:Sct0log}
S\asct{0} = \int d^d x \sqrt{\g} (- \Phi^2 + \frac{1}{8} \Phi \square_\g \Phi + \frac{1}{64} \Phi \square_\g^2 \Phi + \frac{1}{128} r \Phi \square_\g^3 \Phi)\,.
\ee 
Notice that the logarithmic term in \eqref{eq:phirad0} leads to a logarithmic divergence in the bare on-shell action. This divergence has to be cancelled with a logarithmic counterterm, \ie a counterterm featuring an explicit $r$, which is the last term in \eqref{eq:Sct0log}. 

The first variation of the on-shell action takes the form:
\be
\label{eq:deltaSct0log}
\delta (S\as{0} + S\asct{0}) = \int d^2 x (-6 \phi\ab{0}{4} + \frac{1}{96} \square^3 \phi\ab{0}{-2} ) \delta \phi\ab{0}{-2}\,.
\ee 
Notice the prefactor $-6$ equals $\Dm - \Dp$, as in section 2 of \cite{me}, and we also find a non-zero contact term. This contact term is scheme-dependent because one may add to the action an arbitrary multiple of the finite local counterterm given by
\be
\label{eq:Sct0scheme}
\int d^2 x \,\phi\ab{0}{-2} \square^3 \phi\ab{0}{\-2}\,, 
\ee
which is easily seen to modify the coefficient of the contact term in \eqref{eq:deltaSct0log}. 

\paragraph{Conformal anomaly\\}
Let us now consider the derivation of the holographic Callan-Symanzik equation. Although the holographic conformal anomaly for the scalar two-point function is a standard result \cite{deHaro:2000xn,Skenderis:2002wp}, we will derive it here in a slightly different manner which will generalize more easily to the more involved examples in the next subsections.

We start with the simple observation that the renormalized action remains finite as we let the cutoff $r \to \infty$. This in particular implies that its first derivative with respect to $r$ should vanish,
\be
\label{eq:Sinv}
\lim_{r \to \infty} \frac{d}{dr} (S\as{0} + S\asct{0}) = 0\,.
\ee
Now, for every finite value of $r$ the action can be viewed as a functional of the scalar field $\Phi$, the boundary metric $\g_{ij}$ and, in the case of logarithmic counterterms, an explicit function of $r$ itself. This observation allows us to rewrite the total derivative with respect to $r$ as:
\be
\label{eq:bareCSeqn}
\frac{d}{dr} S[\Phi,\g_{ij},r] = \Big(\int d^d x \, \dot \Phi \fdel{\Phi} + \int d^d x \, \dot \g_{ij} \fdel{\g_{ij}} + \frac{\del}{\del r}\Big) S[\Phi,\g_{ij},r] \,.
\ee
In the large $r$ limit we find from \eqref{eq:phirad0} that $\dot \Phi \to 2 \Phi$ and from \eqref{eq:bgmetric} we also obtain that $\dot \g_{ij} = 2 \g_{ij}$. Furthermore, the only term contributing to the partial $r$-derivative is the logarithmic counterterm, so the last counterterm in \eqref{eq:Sct0log}, since it is the only term featuring an explicit $r$-dependence. We may therefore write:
\be
\lim_{r \to \infty} \Big[\Big(\int d^d x \, 2 \Phi \fdel{\Phi} + \int d^d x \, 2 \g_{ij} \fdel{\g_{ij}} \Big) (S\as{0} + S\asct{0}) + \frac{1}{128} \int d^d x \sqrt{\g} \Phi \square^3_\g \Phi \Big] = 0\,.
\ee
The next step is to change variables from $(\Phi,\g_{ij})$ to $(\phi\ab{0}{-2}, g_{ij})$. This is straightforward since as $r \to \infty$ we find that $\Phi \to e^{2r} \phi\ab{0}{-2}$ and similarly  $\g_{ij} \to e^{2r} g_{ij}$ (where $g_{ij} = \delta_{ij}$ in our case).  Furthermore, the combined action $S\as{0} + S\asct{0}$ becomes the renormalized partition function $W\as{0}$ of the dual CFT as $r \to \infty$. We therefore find:
\be
\label{eq:CSeqn}
\Big(\int d^d x \, 2 \phi\ab{0}{-2} \fdel{\phi\ab{0}{-2}} + \int d^d x \, 2 g_{ij} \fdel{g_{ij}} \Big) W\as{0} + \mathcal A[\phi\ab{0}{-2}] = 0\,,
\ee 
with the conformal anomaly given by:
\be
\label{eq:anomalyzerothorder}
\mathcal A[\phi\ab{0}{-2}] = \frac{1}{128} \int d^d x \, \phi\ab{0}{-2} \square^3_0 \phi\ab{0}{-2} \,.
\ee
Equations \eqref{eq:CSeqn} and \eqref{eq:anomalyzerothorder} together give a holographic Callan-Symanzik equation whose functional form matches precisely \eqref{eq:CSmddmW} combined with \eqref{eq:mddmW} and expanded to second order in the sources. Furthermore, the overall coefficient of $1/128$ in \eqref{eq:anomalyzerothorder} also agrees with expectations, since upon comparing with \eqref{eq:CSmddmW} we obtain 
\be
\label{eq:c2holographic}
\frac{1}{128} = \frac{c_2 \pi}{64 \cdot 36} \qquad \rightarrow \qquad c_2 = \frac{18}{\pi}\,.
\ee
Now according to \cite{Freedman:1998tz}, the two-point function obtained from a canonically normalized bulk scalar field has
\be
c_2 = \frac{(2\D - d) \G(\D)}{\pi^{d/2} \G(\D - d/2)}\,,
\ee
which indeed reduces to $18/\pi$ for $d=2$ and $\D = 4$.

\subsection{First correction and three-point function}
\label{sec:firstordercorn}
Let us now work out the first-order correction in $\lambda$. Just as in the previous subsection we will solve the first-order correction to the equation of motion, substitute it into the on-shell action, add counterterms to cancel the divergences and obtain the renormalized one-point function and the holographic Callan-Symanzik equation.  

\paragraph{First-order bulk solution\\}
To find the first-order correction $\Phi\as{1}$ in \eqref{eq:phifullexp} we need to asymptotically solve the second equation in \eqref{eq:eomseries}. The resulting solution takes the form:
\be
\label{eq:expphi1log}
\begin{split}
\Phi\as{1} &= \phi\ab{1}{-4} e^{4r}\\
&\quad + (\tilde \phi\ab{1}{-2} r + 0) e^{2r}\\
&\quad + \tilde \phi\ab{1}{0} r + \phi\ab{1}{0}\\
&\quad + (\tilde \phi\ab{1}{2}r  + \phi\ab{1}{2}) e^{-2r}\\
&\quad + (\tilde{\tilde \phi}\ab{1}{4} r^2 + \tilde \phi\ab{1}{4} r + \phi\ab{1}{4} ) e^{-4r} + \ldots
\end{split}
\ee
Just as in the analogous computation in \cite{me} we find that almost all the coefficients in this expansion can be recursively determined in terms of $\phi\ab{0}{-2}$. The only exceptions are $\phi\ab{1}{4}$, which is not determined in terms of $\phi\ab{0}{-2}$ by the asymptotic analysis alone, and $\phi\ab{1}{-2}$, which is an order $\lambda$ correction to the source and which we set to zero in \eqref{eq:expphi1log}. For the other coefficients we find for example that:
\be
\label{eq:coeffphi1log}
\begin{split}
\phi\ab{1}{-4} &= \frac{\lambda}{16}  \phi\ab{0}{-2}^2\\
\tilde \phi\ab{1}{-2} &= \frac{\lambda}{24} (\phi\ab{0}{-2}\square \phi\ab{0}{-2} - \qt \square \phi\ab{0}{-2}^2)\,.
\end{split} 
\ee
Notice also the presence of the quadratic term $\tilde{\tilde \phi}\ab{1}{4} r^2$ in \eqref{eq:expphi1log}. It arises from the equation of motion in a similar manner as the linear term which we discussed in the previous subsection.

Computing the more subleading coefficients of the various terms in \eqref{eq:expphi1log} becomes an increasingly tedious task. As an example of the complexity involved, let us consider $\tilde{\tilde \phi}\ab{1}{4}$. Using Mathematica we find that it takes the form:
\be
\label{eq:tildetildephi14}
\begin{split}
\tilde{\tilde \phi}\ab{1}{4} = 
\frac{\lambda}{294912}
\Big(
&- \square^4 \phi\ab{0}{-2}^2 + 4 \square^3(\phi\ab{0}{-2} \square \phi\ab{0}{-2}) - 4
\square^3 \phi\ab{0}{-2} \square \phi\ab{0}{-2} \\ & \qquad - 4 \square(\phi\ab{0}{-2} \square^3 \phi\ab{0}{-2}) + 2 \phi\ab{0}{-2} \square^4 \phi\ab{0}{-2}
\Big) \,.
\end{split}
\ee
We emphasize that all steps involved are algebraic so the computer implementation is relatively straightforward.

\paragraph{Counterterm action\\}
The next step is to substitute the solution $\Phi\as{0} + \Phi\as{1}$ into the on-shell action \eqref{eq:sonshell} plus the counterterm action \eqref{eq:Sct0log} which we needed at zeroth order in $\lambda$. We then compute the divergences in the on-shell action which would arise as $r \to \infty$ and find the counterterm action which cancels these divergences. This analysis is again largely analogous to that of section 2 of \cite{me}, in particular one again obtains the so-called `pseudo-non-local' divergences discussed in \cite{me} which are again all cancelled by local counterterms.

Rather than writing down the full form of the counterterm action, which is rather involved, let us present here only the logarithmic terms which are relevant for the computation of the anomaly below. These take the form:
\be
\label{eq:Sct1log}
\begin{split}
S\asct{1} &= \lambda \int d^d x \sqrt{\g} \Big( \ldots - \frac{r}{1024} \Phi^2 \square_\g^3 \Phi + \frac{r^2}{12288} \big[\Phi^2 \square^4_\g \Phi - 4 \Phi \square^3_\g(\Phi\square_\g\Phi)\big]
\\& \qquad + r\big[\frac{9223}{73728}\Phi^2 \square^4_\g \Phi - \frac{1}{12288} \Phi \square_\g^3 (\Phi \square_\g \Phi) - \frac{1}{4096} \Phi \square^2_\g \Phi \square_\g^2 \Phi \big]
+
\ldots\Big)\,.
\end{split}
\ee
Notice the appearance of the term involving an explicit $r^2$ which corresponds to a divergence of the form $\log^2(\Lambda/\mu)$ in the field theory.

\paragraph{One-point function\\}
The one-point function is again obtained by following the same procedure as in \cite{me}, so in particular we again need to take into account the more leading terms in \eqref{eq:expphi1log} which complicate the relation between $\Phi$ and the source $\phi\ab{0}{-2}$. 
This can be done straightforwardly, see again \cite{me}, and eventually we find the first variation of the on-shell action to be:
\be
\label{eq:firstorderoneptfuntction}
\delta (S + S\asct{0} + S\asct{1}) =  \int d^2 x (-6 [\phi\ab{0}{4} + \phi\ab{1}{4}] + (\text{contact terms}) ) \delta \phi\ab{0}{-2}\,.
\ee
The term in square brackets is the non-local term which is exactly of the expected form and matches the result at order $\lambda$ of \cite{me}. The contact terms at order $\lambda$ involve both local contact terms, for example the term
\be
\lambda \square \phi\ab{0}{-2} \phi\ab{0}{4}\,,
\ee 
as well as ultralocal terms which are for example of the form
\be
\lambda \square^4 \phi\ab{0}{-2} \phi\ab{0}{-2}\,.
\ee
The number of boxes in these expressions is fixed by demanding that these contact terms have the correct conformal weight. Notice that the boxes may act in different ways on the sources which leads to a proliferation of terms. The exact form of the contact terms is however again scheme-dependent.

\paragraph{Conformal anomaly\\}
Let us now investigate the conformal anomaly for the three-point function. Just as in the previous subsection, our starting point is the observation that the radial derivative of the renormalized on-shell action should vanish, which we may write in terms of functional derivatives as 
\be
\label{eq:CSeqn1}
0 = \lim_{r \to \infty} \Big(\int d^d x \, \dot \Phi \fdel{\Phi} + \int d^d x \, \dot \g_{ij} \fdel{\g_{ij}} + \frac{\del}{\del r}\Big) (S + S\asct{0} + S\asct{1})\,.
\ee 
We should convert this into an equation for the renormalized partition function by changing variables from $\Phi$ to $\phi\ab{0}{-2}$ and from $\g_{ij}$ to the boundary metric $g_{ij}$. Although the change of variables for the metric is trivial, this is not the case for the scalar field itself because of the leading terms in \eqref{eq:expphi1log}. Indeed, we find from \eqref{eq:expphi1log} and \eqref{eq:coeffphi1log} that:
\be
\label{eq:leadingexpPhi1log}
\Phi = \frac{\lambda}{16} e^{4r} \phi\ab{0}{-2}^2 + \frac{\lambda}{24} r e^{2r} [\phi\ab{0}{-2}\square \phi\ab{0}{-2} - \qt \square_0 \phi\ab{0}{-2}^2] + e^{2r} \phi\ab{0}{-2} + \ldots
\ee
Rather than working out the change of variables explicitly, let us employ the following trick. We rewrite schematically:
\be
\label{eq:rewritefnder}
\begin{split}
\int \, \dot \Phi \fdel{\Phi} + \int \, \dot \g_{ij} \fdel{\g_{ij}} =
 \int \, \Big(\int \, \dot \Phi \fdel{\Phi} + \int \, \dot \g_{ij} \fdel{\g_{ij}}\Big) \phi\ab{0}{-2} \fdel{\phi\ab{0}{-2}} + 2 \int g_{ij} \fdel{g_{ij}} \,,
\end{split}
\ee
where we used that $\g_{ij} = e^{2r} g_{ij}$ and therefore $\dot \g_{ij} = 2 \g_{ij}$ in our case to obtain the last term.  In the next subsection we will furthermore explain that the first term is given by:
\be
\label{eq:functdersource}
\begin{split}
& \Big(\int d^2 x \, \dot \Phi \fdel{\Phi} + \int d^2 x\, \dot \g_{ij} \fdel{\g_{ij}}\Big) \phi\ab{0}{-2} = 2 \phi\ab{0}{-2}  + \tilde \phi\ab{1}{-2} \\& \qquad = 2 \phi\ab{0}{-2} + \frac{\lambda}{24} [\phi\ab{0}{-2} \square_0 \phi\ab{0}{-2} - \qt \square_0 \phi\ab{0}{-2}^2] + \ldots
\end{split} 
\ee
where $\tilde \phi\ab{1}{-2}$ is the term of order $\exp(2 r) r$ in the radial expansion of $\Phi\as{1}$ and was given in \eqref{eq:coeffphi1log}. We now substitute this result into \eqref{eq:rewritefnder} and then substitute everything back into \eqref{eq:CSeqn1} to obtain the Callan-Symanzik equation to first order in $\lambda$:
\be
\label{eq:CSeqnfirstorder}
\begin{split}
&\int d^d x \Big( 2 \phi\ab{0}{-2} + \frac{\lambda}{24} [\phi\ab{0}{-2} \square_0 \phi\ab{0}{-2} - \qt \square_0 \phi\ab{0}{-2}^2]\Big) \fdel[W]{\phi\ab{0}{-2}} \\& \qquad + \int d^d x \, 2 g_{ij} \fdel[W]{g_{ij}} + \mathcal A[\phi\ab{0}{-2}] = 0\,,
\end{split}
\ee
where the conformal anomaly $\mathcal A$ is defined as:
\be
\label{eq:anomalyfirstorder}
\mathcal A [\phi\ab{0}{-2}] = \lim_{r \to \infty} \frac{\del}{\del r} (S\asct{0} + S\asct{1})\,.
\ee
The logarithmic terms in the counterterm action were given in \eqref{eq:Sct0log} and \eqref{eq:Sct1log}. From their explicit expression it is not obvious that the expression on the right-hand side of \eqref{eq:anomalyfirstorder} is finite. However, this must be the case since all other terms in \eqref{eq:CSeqnfirstorder} are finite. Indeed, using Mathematica we find $\mathcal A$ to be a finite local function of $\phi\ab{0}{-2}$ with the precise form:
\be
\label{eq:preciseanomaly3pt}
\begin{split}
\mathcal A &= 
\int d^2 x \Big[\frac{1}{128} \phi\ab{0}{-2} \square_0^3 \phi\ab{0}{-2}\\ & \qquad \qquad + 
\frac{\lambda}{12288}\Big( \phi \ab{0}{-2}{}^2 \square_0^4 \phi \ab{0}{-2} -4 \phi\ab{0}{-2} \square_0^3  \phi \ab{0}{-2} \square_0 \phi \ab{0}{-2} \Big) \\ & \qquad \qquad - \frac{\lambda}{4096} \Big( \square_0^2 \phi \ab{0}{-2} \left(\square_0 \phi \ab{0}{-2}\right){}^2 +    \left(\square_0^2 \phi \ab{0}{-2}\right)^2 \phi \ab{0}{-2}   \Big)\Big]\,.
\end{split}
\ee  
Notice that we already obtained the leading term in \eqref{eq:anomalyzerothorder}. The terms on the second line turn out to be scheme-dependent, since their overall coefficient can be shifted by adding a finite counterterm at the level of the two-point function as we explained at the end of subsection \ref{sec:2ptfn}. Finally, the terms on the third line are independent of this choice for the renormalization scheme. As far as we are aware the expression \eqref{eq:preciseanomaly3pt} for the conformal anomaly of a scalar three-point function is a new result.

Upon comparing \eqref{eq:CSeqnfirstorder} with \eqref{eq:CSmddmW} and using \eqref{eq:mddmW} we see that the functional form of \eqref{eq:CSeqnfirstorder} precisely matches CFT expectations. From these equations we also obtain a prediction for the coefficient of the three-point function:
\be
\frac{\l}{24} = \frac{\pi c_3}{4 c_2} \qquad \rightarrow \qquad c_3 = \frac{\lambda c_2}{6 \pi} = \frac{3 \lambda}{\pi^2}\,,
\ee
where we used \eqref{eq:c2holographic}. Now, the correlation function of three identical scalar operators is given by $2\lambda$ times the three-point scalar Witten diagram, where an extra factor of $2$ arises from the functional differentation of the one-point function which is quadratic in the source. The overall coefficient obtained from the Witten diagram was presented in equation (25) of \cite{Freedman:1998tz} and for identical operator dimensions results in:
\be
c_3 = - (2 \lambda) \frac{\G(\D/2)^3 \G((3\D - d)/2)}{2 \pi^d \G(\D - d/2)^3}\,,
\ee
which indeed precisely reduces to $3 \lambda/\pi^2$ for $d = 2$ and $\D = 4$.

To summarize, we found that both the functional form and the overall coefficient of the anomaly match field theory expectations. It is amusing to note that we were able to compute the normalization of the three-point function without performing a complete bulk analysis.

\subsection{Extension to higher orders}
\label{sec:higherorderst}
Let us now explain the relation \eqref{eq:functdersource}. To this end we notice that the  operator
\be
\delta_r \equiv \int \dot \Phi \fdel{\Phi} + \int \dot \g_{ij} \fdel{\g_{ij}}
\ee
implements precisely an infinitesimal diffeomorphism given by a constant shift in the radial direction. Since the equations of motion are diffeomorphism invariant, this operator maps solutions to solutions. Furthermore, for any solution to the equation of motion the term $\phi\ab{0}{-2}$ is precisely the coefficient of the term $\exp(2 r)$ in the radial expansion of $\Phi$. Under a small shift in $r$ this term is straightforwardly found to change by:
\be
\delta_r \phi\ab{0}{-2} = 2 \phi\ab{0}{-2} + \tilde \phi\ab{1}{-2}
\ee
and this explains \eqref{eq:functdersource}. 

The above argument however directly extends to all orders in $\lambda$, at least as long as we do not have multi-trace counterterms. Indeed, we may use exactly the same reasoning to write:
\be
\delta_r \phi\ab{0}{-2} = 2 \phi\ab{0}{-2} + \tilde \phi\bs{-2}\,,
\ee
where now the term $\tilde \phi\bs{-2}$ on the right-hand side is the term of order $\exp(2 r) r$ in the radial expansion of $\Phi$ to any given order in $\lambda$. In the absence of multi-trace counterterms the all-order holographic Callan-Symanzik equation can therefore be compactly written as:
\be
\label{eq:generalholCSstS}
\Big( \int  (2 \phi\ab{0}{-2} + \tilde \phi\bs{-2})  \fdel{\phi\ab{0}{-2}} + 2 \int  g_{ij} \fdel{g_{ij}} \Big) W + 
\lim_{r \to \infty} \frac{\del}{\del r} S_{\text{ct}} = 0\,.
\ee

Furthermore, this method also direcly extends to the normalizable mode $\phi\bs{4}$. In that case the corresponding result is given by
\be
\delta_r \phi\bs{4} = - 4 \phi\bs{4} + \tilde \phi\bs{4}\,,
\ee
which again holds to all orders in $\lambda$. We may now repeat the above change of variables from $(\Phi,\g_{ij})$ to $(\phi\ab{0}{-2},g_{ij})$ to arrive at the following all-order Callan-Symanzik equation at the level of the one-point function:
\be
\label{eq:generalholCSst}
\Big( \int (2 \phi\ab{0}{-2} + \tilde \phi\bs{-2}) \fdel{\phi\ab{0}{-2}} + 2 \int g_{ij} \fdel{g_{ij}} \Big) \phi\bs{4}(x) = - 4 \phi\bs{4}(x) + \tilde \phi\bs{4} (x)\,.
\ee
Notice that this equation is completely equivalent to \eqref{eq:generalholCSstS} and gives an alternative way to compute the (first variation of the) conformal anomaly $\mathcal A$.

Let us emphasize again that the equations \eqref{eq:generalholCSstS} and \eqref{eq:generalholCSst} hold up to any order in $\lambda$ \emph{provided} we do not need any multi-trace counterterms. The necessary modifications when multi-trace counterterms do appear are discussed below. 

We note that obtaining conformal anomalies by exploiting the asymptotic equivalence between radial diffeomorphisms and scale tranformations is a standard result which dates back to the early days of AdS/CFT \cite{Henningson:1998ey,Henningson:1998gx}. Nevertheless, the above extension to irrelevant operators, which in particular includes more leading terms in the radial expansion and corresponding nontrivial beta functions, appears not to have been considered so far. 

Finally, it is of course well-known that the above reasoning can also be extended to incorporate \emph{local} rescalings, \ie Weyl transformations. These are generated by specific bulk diffeomorphisms called \emph{PBH transformations} and discussed in for example \cite{Imbimbo:1999bj,Schwimmer:2000cu,Skenderis:2000in,Erdmenger:2001ja}. For these transformations the corresponding conformal Ward identity can be obtained by replacing \eqref{eq:Sinv} by invariance under such a general PBH transformation which follows direcly from diffeomorphism invariance of the bare on-shell action (\ie from the ordinary constraints of general relativity). It would be interesting to generalize the results presented here to include such local Weyl rescalings.

\subsection{Third-order correction and multi-trace counterterms}
\label{sec:thirdorder}
At second and third order in $\lambda$ the computation of the counterterms parallels the discussion in section 2 of \cite{me}, plus additional logarithmic divergences which we may deal with in the same manner as in the previous subsection. Rather than performing a full analysis we will in this subsection focus on the novelties brought about by the \emph{multi-trace} counterterm. As explained in \cite{me}, such a term arises at order $\lambda^3$ if the scaling dimension $\D \geq 2d$. Since in our case $\D = 2d$ we find by power counting that this counterterm should become of a logarithmic form. This in turn should lead to mixing between single- and multi-trace operators and below we will see that this indeed turns out to be the case.   

In the following we omit all the terms involving the box $\square$. This will allow us to focus on the multi-trace divergence and counterterm without having to deal with the added complexity of these subleading terms.

\paragraph{Third-order bulk solution\\}
The solution to the equation of motion up to order $\lambda^3$ takes the following form:
\bea
\Phi &=& e^{2r} \phi\ab{0}{-2} + e^{-4r} \phi\ab{0}{4} + \ldots\nonumber\\
&&+ \lambda \Big( \frac{1}{16} e^{4r} \phi\ab{0}{-2}^2 - \qt e^{-2r} \phi\ab{0}{-2}\phi\ab{0}{4} + e^{-4r} \phi\ab{1}{4} + \ldots \Big) \nonumber\\
&&+ \lambda^2 \Big( \frac{1}{320} e^{6r} \phi\ab{0}{-2}^3 + \frac{3}{64} \phi\ab{0}{-2}^2 \phi\ab{0}{4} + e^{-4r} \phi\ab{2}{4} +\ldots \Big) \label{eq:philambda3}\\
&&+ \lambda^3 \Big(\frac{13}{92160} e^{8r}\phi\ab{0}{-2}^4 + \frac{11}{960} e^{2r} r \phi\ab{0}{-2}^3 \phi\ab{0}{4} \nonumber\\ &&\qquad \qquad - \frac{11}{640} e^{-4r} r \phi\ab{0}{-2}^2 \phi\ab{0}{4} + e^{-4r} \phi\ab{3}{4} + \ldots  \Big)\,, \nonumber
\eea
where here and below the dots represent terms involving boxes. Notice the appearance of the term of order $\exp(2r) r$ at order $\lambda^3$ which involves the non-locally determined term $\phi\ab{0}{4}$ but is \emph{more} leading than the source term of order $\exp(2r)$ on the first line. As explained in \cite{me} this is a sign that the variational principle will have to be modified and therefore multi-trace counterterms should appear.

\paragraph{Counterterm action\\}
It is straightforward to find that the counterterm action takes the form:
\be
\label{eq:Scttriple}
S_{\text{ct}} = \int d^2 x \sqrt{\g} \Big( - \Phi^2 - \frac{\lambda}{24}\Phi^3 + \frac{\lambda^2}{1280}\Phi ^4 - \frac{\lambda ^3}{30720}\Phi^5 + \frac{11 \lambda^3}{11520}r \Phi^3 \Pi_r^2
+ \ldots\Big)\,,
\ee
again up to terms involving boxes. The last counterterm involves the \emph{renormalized conjugate momentum} $\Pi_r$ which is defined as:
\be
\label{eq:pir}
\Pi_r = \dot \Phi - 2 \Phi + \ldots = -6 \phi\ab{0}{4} e^{-4r} + \ldots
\ee
In \cite{me} it is explained that counterterms involving the conjugate momentum should be interpreted as multi-trace counterterms. As  announced in the introduction to this subsection, in the present case the multi-trace counterterm indeed arises at order $\lambda^3$ and is of a logarithmic form.

As in section \ref{sec:freefield} we again find a scheme dependence associated to the logarithmic divergence. Here it arises by considering the particular counterterm
\be
\label{eq:Scttilde}
\begin{split}
\tilde S_{\text{ct}} &= a \int d^2 x \sqrt{\g} \frac{11 \lambda^3}{11520} \Phi^3 \Pi_r^2 = a \int d^2 x \frac{11 \lambda^3}{320} \phi\ab{0}{-2}^3 \phi\ab{0}{4}^2 + \ldots
\end{split}
\ee
with an arbitrary finite real coefficient $a$. (We inserted the extra $11/11520$ for later convenience.) Since this term is finite as $r \to \infty$ we may freely add it to the on-shell action and changing $a$ then corresponds to a change in the renormalization scheme. Since we would like to make this scheme explicit dependence below, we continue to include this counterterm with arbitrary $a$.

\paragraph{One-point function\\}
The total variation of the on-shell action plus the counterterm action is given by:
\be
\label{eq:deltaSmultitrace}
\begin{split}
&\delta (S + S_\text{ct} + \tilde S_{\text{ct}}) = \int d^2 x \sqrt{\g}\, \dot \Phi \delta \Phi + \delta S_\text{ct} + \delta \tilde S_\text{ct} = \\& \int d^2 x \sqrt{\g} \,
\Big(\Pi_r  - \frac{\lambda}{8}\Phi^2 + \frac{\lambda^2}{320}\Phi^3 - \frac{\lambda^3}{6144}\Phi^4 + \frac{11 \lambda^3}{3840}(r+a) \Phi^2 \Pi_r^2 + \ldots \Big) \delta \Phi \\ & \qquad+ \int d^2 x \sqrt{\g} \, \frac{11 \lambda^3}{5760} (r+a) \Phi^3 \Pi_r \delta \Pi_r 
\end{split}
\ee
and in terms of the variables $\phi\ab{0}{-2}$ and $\phi\ab{0}{4}$ we find that this variation becomes:
\be
\label{eq:deltaS3}
\begin{split}
&\delta(S + S_\text{ct} + \tilde S_\text{ct}) =\\ &\qquad \int d^2 x \Big( -6 \sum_{i=0}^4\phi \ab{i}{4} + \lambda^3 (\frac{993}{12800}  + \frac{33}{320} a ) \phi\ab{0}{-2}^2 \phi\ab{0}{4}^2 + \ldots \Big) \delta \phi\ab{0}{-2}\\
&\qquad + \int d^2 x \lambda^3 (\frac{35}{768} + \frac{11}{160}a) \phi\ab{0}{-2}^3 \phi\ab{0}{4} \delta \phi\ab{0}{4} \,.
\end{split} 
\ee
The term multiplying $\delta \phi\ab{0}{-2}$ is the expected non-local contribution to the one-point function plus a local scheme-dependent contact term. However on the last line we find an unusual additional term proportional to the variation $\delta \phi\ab{0}{4}$. The appearance of this term implies that the proper boundary data is no longer $\phi\ab{0}{-2}$ but rather $\chi \bs{-2}$ which is defined as:
\be
\label{eq:tildephiab0dm}
\phi\ab{0}{-2} = \chi\bs{-2} + \lambda^3 \frac{1}{6} (\frac{35}{768} + \frac{11}{160} a) \chi\bs{-2}^3 \phi\ab{0}{4}\,.
\ee
Indeed, the change of variables from $\phi\ab{0}{-2}$ to $\chi\bs{-2}$ leads to a proper variation of the on-shell action of the form:
\be
\label{eq:deltaSshiftedsource}
\begin{split}
&\delta(S + S_\text{ct} + \tilde S_\text{ct}) =\\ &\qquad \int d^2 x \Big( -6 \sum_{i=0}^4\phi \ab{i}{4} - \lambda^3 ( \frac{757}{12800} + \frac{33}{320} a ) \chi\bs{-2}^2 \phi\ab{0}{4}^2 + \ldots \Big) \delta \chi\bs{-2}\,.
\end{split}
\ee
Following the usual multi-trace literature \cite{Berkooz:2002ug,Witten:2001ua,Mueck:2002gm,Sever:2002fk,Elitzur:2005kz,Papadimitriou:2007sj} the extra variation involving $\delta \phi\ab{0}{4}$ and subsequent redefinition of the sources can be interpreted as an insertion of the multi-trace operator $\op^2$. In our case this insertion is induced by the counterterms and arises from mixing between the single- and the multi-trace operator. Consequently, its precise form should be scheme-dependent and this is indeed reflected by the explicit $a$ appearing in \eqref{eq:deltaS3}. Notice that such mixing could not occur in \cite{me} because operator mixing only arises in the presence of logarithmic divergences.

\paragraph{Conformal anomaly\\}
We may now find the holographic Callan-Symanzik equation as follows. As before, we begin with:
\be
\lim_{r \to \infty} \frac{d}{dr} (S+S_\text{ct} + \tilde S_\text{ct}) = 0
\ee
and then rewrite the radial derivative as a functional derivative. From \eqref{eq:deltaSmultitrace} we see that it is natural to regard the on-shell action as a function of both $\Phi$ and $\Pi_r$. We may therefore write:
\be
\lim_{r \to \infty} \Big(\int \dot \Phi \fdel{\Phi} + \int \dot \Pi_r \fdel{\Pi_r} + \int \dot \g_{ij} \fdel{\g_{ij}} + \frac{\del}{\del r} \Big)  (S+S_\text{ct} + \tilde S_\text{ct}) = 0\,.
\ee
The next logical step would now be to change variables to $\phi\ab{0}{-2}$ and $\phi\ab{0}{4}$, which we may similarly regard as independent variables, and then finally change variables to $\chi\bs{-2}$. It is however easier to work in a scheme where $\chi\bs{-2} = \phi\ab{0}{-2}$. According to \eqref{eq:tildephiab0dm} this can be done by choosing
\be
a = - \frac{35}{768} \frac{160}{11}\,.
\ee 
We notice that a shift in $a$ does not introduce extra scale dependence so the choice for this scheme will not affect the final form of the Callan-Symanzik equation. In that case we may write again:
\be
\begin{split}
&\int \, \dot \Phi \fdel{\Phi} +  \int \, \dot \Pi_r \fdel{\Pi_r} + \int \, \dot \g_{ij} \fdel{\g_{ij}}  =\\
&\int \, \Big(\int \, \dot \Phi \fdel{\Phi} +  \int \, \dot \Pi_r \fdel{\Pi_r} + \int \, \dot \g_{ij} \fdel{\g_{ij}}\Big) \phi\ab{0}{-2} \fdel{\phi\ab{0}{-2}} + 2 \int g_{ij} \fdel{g_{ij}}= \\
&\int \, (2 \phi\ab{0}{-2} + \tilde \phi\ab{0}{-2} ) \fdel{\phi\ab{0}{-2}} + 2 \int g_{ij} \fdel{g_{ij}}\,,
\end{split}
\ee
where we arrived at the last line by using again the reasoning of subsection \ref{sec:higherorderst}. From \eqref{eq:philambda3} we may read off that:
\be
\tilde \phi\ab{0}{-2} = \frac{11}{960} \lambda^3 \phi\ab{0}{-2}\phi\ab{0}{4}
\ee
and $\del_r S_{\text{ct}}$ is just given by \eqref{eq:Scttilde} with $a=1$. We therefore find that
\be
\int (2 \phi\ab{0}{-2} + \frac{11}{960} \lambda^3 \phi\ab{0}{-2} \phi\ab{0}{4} + \ldots) \fdel[W]{\phi\ab{0}{-2}} + \int \frac{11}{320} \lambda^3 \phi\ab{0}{-2}^2 \phi\ab{0}{4}^2 = 0\,,
\ee
which we may rewrite as:
\be
\int (2  \phi\ab{0}{-2} + \ldots) \fdel[W]{\phi\ab{0}{-2}} + \int 2 g_{ij} \fdel[W]{g_{ij}}- \int \frac{11 \lambda^3}{11520} \phi\ab{0}{-2}^3 \Big(\fdel[W]{\phi\ab{0}{-2}}\Big)^2 = 0\,.
\ee
As we have just argued, in this equation one may directly replace $\phi\ab{0}{-2}$ with $\chi\bs{-2}$ when one works in a different renormalization scheme. The final Callan-Symanzik equation is therefore:
\be
\int (2 \chi\bs{-2} + \ldots) \fdel[W]{\chi\bs{-2}} + \int 2 g_{ij} \fdel[W]{g_{ij}}- \int \frac{11 \lambda^3}{11520} \chi\bs{-2}^3 \Big(\fdel[W]{\chi\bs{-2}}\Big)^2 = 0\,.
\ee
This equation is precisely of the form \eqref{eq:nonlincs} and therefore again matches the field theory expectations. 

\subsection{Multi-trace counterterms at higher orders}
Let us now discuss the systematics of holographic renormalization in the presence of multi-trace counterterms at higher orders in $\lambda$. As in the previous subsection we will continue to omit terms involving the box $\square$. Within this context we will work up to terms of order $\lambda^7$, which will allow us to also incorporate the effects of a triple-trace counterterm.

We note that the results below rely heavily on Mathematica and the methods we will discuss are correspondingly best suited for a computer implementation. Furthermore, to avoid clutter we will henceforth omit the dots which indicated terms involving boxes in the previous subsection.  

\paragraph{Seventh-order bulk solution\\}
At order $\lambda^7$ the bulk solution as expected becomes rather involved. The terms relevant to the discussion below take the following form:
\be
\Phi = \ldots + e^{2r}r \tilde \phi\bs{-2} + e^{2r} \phi\ab{0}{-2} + \ldots + e^{-4r} r \tilde \phi\bs{4} + \phi\bs{4} + \ldots 
\ee
where $\phi\ab{0}{-2}$ and $\phi\bs{4}$ are left undetermined by the radial expansion and: 
\be
\label{eq:tildephibsmtf}
\begin{split}
\tilde \phi\bs{-2} &= \lambda^3 \frac{11}{960} \phi\ab{0}{-2}^3 \phi\bs{4} - \lambda^6 \frac{17413}{86016000} \phi\ab{0}{-2}^5 \phi\bs{4}^2 \\
\tilde \phi\bs{4} &=  - \lambda^3 \frac{11}{640} \phi\ab{0}{-2}^2 \phi\bs{4}^2 + \lambda^6 \frac{132869}{387072000}\phi\ab{0}{-2}^4 \phi\bs{4}^3 \,,
\end{split} 
\ee
where we should remember that $\phi\bs{4}$ also has an expansion in $\lambda$ but $\phi\ab{0}{-2}$ does not since we continue to pick a bulk solution such that this coefficient is fixed at all orders. Notice also that the numerical coefficients appearing in \eqref{eq:tildephibsmtf} and below are specific to the toy model at hand and therefore do not carry much physical significance.

\paragraph{Counterterm action\\}
The first noteworthy aspect of the holographic renormalization beyond order $\lambda^3$ is the fact that the conjugate momentum $\Pi_r$ also has an expansion in $\lambda$. This we may incoporate as follows.

At any order $\lambda^k$ the most leading divergence can be cancelled with a counterterm of the form $\lambda^k \Phi^{k+2}$. For example, up to order $\lambda^7$ we find that these `single-trace' terms have the following form:
\be
\begin{split}
S^{\text{st}}_{\text{ct}} &= \int d^2 x \sqrt{\g} \Big(- \Phi ^2 - \frac{\lambda }{24}\Phi ^3 + \frac{\lambda ^2}{1280}\Phi ^4 - \frac{\lambda ^3}{30720}\Phi ^5 \\ & \qquad + \frac{31 \lambda ^4}{17203200}\Phi ^6 - \frac{\lambda ^5}{8601600}\Phi ^7 + \frac{9827 \lambda ^6}{1189085184000}\Phi ^8 - \frac{29887\lambda ^7 }{47563407360000} \Phi ^9\Big)\,.
\end{split}
\ee
Notice that this matches with the single-trace part of \eqref{eq:Scttriple} up to order $\lambda^3$. After adding these counterterms we define the $\lambda$-corrected renormalized conjugate momentum as:
\be
\Pi_r^\lambda = \dot \Phi + \frac{1}{\sqrt \g} \frac{\delta S^{\text{st}}_{\text{ct}}}{\delta \Phi}
\ee
For $\lambda = 0$, this definition agrees with the renormalized conjugate momentum $\Pi_r$ as defined above but at higher orders it is manifestly different. Up to order $\lambda^4$ we trivially find:
\be
\Pi_r^\lambda = \dot \Phi  - 2 \Phi - \frac{\lambda}{8}\Phi^2 + \frac{4 \lambda ^2}{1280}\Phi^3 - \frac{5 \lambda ^3}{30720}\Phi^4 + \frac{186 \lambda ^4}{17203200}\Phi ^5 \,.
\ee
We will not need any higher-order corrections to $\Pi_r^\lambda$ below.

The next step is to add the multi-trace counterterm action. At order $\lambda^3$ we already found the term quadratic in $\Pi_r^\lambda$ which is the last term \eqref{eq:Scttriple}. At higher orders we find that the term quadratic in $\Pi_r^\lambda$ becomes:
\be
\begin{split}
S^{\text{dt}}_{\text{ct}} &= \frac{11}{11520} \int d^2 x \sqrt{\g} (\Pi_r^\lambda)^2 \Big[  r \lambda^3 \Phi^3 + \left(\frac{1}{16} r - \frac{1157}{49280} \right) \lambda^4 \Phi^4  \\ &\qquad -\left(\frac{3}{1280}r - \frac{3321}{7884800}\right) \lambda^5\Phi^5 +\left( \frac{11}{92160} r + \frac{19867}{10218700800} \right)\lambda^6 \Phi^6 \\&\qquad - \left(\frac{53}{7372800} r + \frac{440057}{286123622400}\right)\lambda^7 \Phi^7\Big]
\end{split} 
\ee
and we interpret all of these as double-trace divergences in the dual field theory. Starting at order $\lambda^6$ we find that we furthermore need a cubic term:
\be
S^\text{tt}_\text{ct} = \frac{2981}{870912000} \int d^2 x \sqrt{\g} (\Pi_r^\lambda)^3 \Big[  r \lambda^6 \Phi^5 + \left(\frac{20815}{277504} r - \frac{3763891}{146522112}\right)\lambda^7\Phi^6 \Big]\,.
\ee
These terms are interpreted as triple-trace counterterms. The fact that such terms first appear as a logarithmic divergence at order $\lambda^6$ is easily verified to match expectations from power counting in the dual field theory.

Notice that we never needed a counterterm which is \emph{linear} in $\Pi_r^\lambda$ at least to order $\lambda^7$. As in \cite{me}, we claim that the absence of such counterterms is systematic and that any divergences that would require such counterterms are in fact pseudo-non-local.

\paragraph{One-point function\\}
The first variation of the on-shell action now takes the form:
\begin{multline}
\label{eq:deltaSmultimulti}
\delta (S + S_\text{ct}^\text{st} + S_\text{ct}^\text{dt} + S_\text{ct}^\text{tt}) = \\  \int d^2 x \Big( - 6 \phi\bs{4} + \lambda^3 \frac{993}{12800} \phi\ab{0}{-2}^2 \phi\bs{4}^2 - \lambda^6\frac{12226986017}{2601123840000} \phi\ab{0}{-2}^4 \phi\bs{4}^3 \Big) \delta \phi\ab{0}{-2} \\ \qquad + \int d^2 x \Big( \lambda^3 \frac{35}{768} \phi\ab{0}{-2}^3 \phi\bs{4} - \lambda^6 \frac{309141727}{123863040000} \phi\ab{0}{-2}^5 \phi\bs{4}^2\Big) \delta \phi\bs{4}\,, 
\end{multline}
where we already obtained the result at order $\lambda^3$ in \eqref{eq:deltaS3}. The proper source $\chi\bs{-2}$ is to this order defined via:
\be
\label{eq:chibs}
\phi\ab{0}{-2} = \chi\bs{-2} + \lambda^3 \frac{35}{4608} \chi\bs{-2}^3 \phi\bs{4}- \lambda^6 \frac{107531227}{1486356480000} \chi\bs{-2}^5 \phi\bs{4}^2\,,
\ee
since this redefinition ensures that the variation of the on-shell action becomes:
\begin{multline}
\label{eq:deltaSmmchi}
\delta (S + S_\text{ct}^\text{st} + S_\text{ct}^\text{dt} + S_\text{ct}^\text{tt}) = \\  \int d^2 x \Big( - 6 \phi\bs{4} - \lambda^3 \frac{757}{12800} \chi\bs{-2}^2 \phi\bs{4}^2 - \lambda^6\frac{2163761801}{5202247680000} \chi\bs{-2}^4 \phi\bs{4}^3 \Big) \delta \chi \bs{-2}\,.
\end{multline}
The redefinition \eqref{eq:chibs} implies that operator mixing occurs between single-, double- and triple-trace counterterms.

\paragraph{Conformal anomalies\\}
As before, the holographic Callan-Symanzik will be obtained from:
\be
\label{eq:bareCSeqnmultimultitrace}
\lim_{r \to \infty} \Big(\int \dot \Phi \fdel{\Phi} + \int \dot \Pi_r \fdel{\Pi_r} + \int \dot \g_{ij} \fdel{\g_{ij}} + \frac{\del}{\del r} \Big)  (S+ S_\text{ct}^\text{st} + S_\text{ct}^\text{dt} + S_\text{ct}^\text{tt} )= 0\,.
\ee
Upon changing variables to $\phi\ab{0}{-2}$, $\phi\bs{4}$ and $g_{ij}$ we find that:
\be
\label{eq:delrmultimultitrace}
\begin{split}
&\int \dot \Phi \fdel{\Phi} + \int \dot \Pi_r \fdel{\Pi_r} + \int \dot \g_{ij} \fdel{\g_{ij}} =\\ 
&\qquad \int ( 2 \phi\ab{0}{-2} + \tilde \phi\bs{-2}) \fdel{\phi\ab{0}{-2}} + \int (- 4 \phi\bs{4} + \tilde \phi\bs{4}) \fdel{\phi\bs{4}} + \int 2 g_{ij} \fdel{g_{ij}}\,,
\end{split}  
\ee
where we used the same reasoning as in subsection \ref{sec:higherorderst}. The coefficients $\tilde \phi\bs{-2}$ and $\tilde \phi\bs{4}$ were given in \eqref{eq:tildephibsmtf}. Furthermore, the partial $r$-derivative of the counterterm action is given by:
\be
\label{eq:confanomalyhigherorder}
\lim_{r \to \infty} \del_r (S_\text{ct}^\text{st} + S_\text{ct}^\text{dt} + S_\text{ct}^\text{tt}) = \int d^2 x \Big(\lambda^3 \frac{11}{320} \phi\ab{0}{-2}^2 \phi\bs{4}^2 - \lambda^6 \frac{109021}{129024000} \phi\ab{0}{-2}^5 \phi\bs{4}^3 \Big)\,.
\ee
We now subsitute \eqref{eq:tildephibsmtf} into \eqref{eq:delrmultimultitrace} and then substitute this back into \eqref{eq:bareCSeqnmultimultitrace}. Substituting then also \eqref{eq:confanomalyhigherorder} and \eqref{eq:deltaSmultimulti} for the variation of the action with respect to $\phi\ab{0}{-2}$ and $\phi\bs{4}$ we arrive at:
\be
\begin{split}
&\int d^2 x \Big(- 12 \phi\ab{0}{-2} \phi\bs{4} - \lambda^3 \frac{1181}{19200} \phi\ab{0}{-2}^3 \phi\bs{4}^2 + \\&\qquad \lambda^6 \frac{1375122517}{1300561920000} \phi\ab{0}{-2}^5 \phi\bs{4}^3 \Big) + \int g_{ij} \fdel{g_{ij}} W = 0\,.
\end{split}
\ee
We may now finally change variables to $\chi\bs{-2}$ using \eqref{eq:chibs} and also iteratively use \eqref{eq:deltaSmmchi} to rewrite $\phi\bs{4}$ in terms of $\delta W/ \delta \chi\bs{-2}$. This ultimately leads to:
\be
\begin{split}
&\int 2 \chi\bs{-2} \fdel[W]{\chi\bs{-2}} + \int 2 g_{ij} \fdel[W]{2 g_{ij}}- \lambda^3 \int \frac{11}{11520} \chi\bs{-2}^3 \Big(\fdel[W]{\chi\bs{-2}}\Big)^2  \\ & \qquad - \lambda^6 \int \frac{2981}{1741824000}\chi\bs{-2}^5 \Big(\fdel[W]{\chi\bs{-2}}\Big)^3 = 0\,.
\end{split}
\ee
As expected, we indeed find operator mixing at the triple-trace level. We may straightforwardly read off the beta functions for the double- and triple-trace operators in the presence of the source $\chi\bs{-2}$ for the single-trace operator to be:
\be
\begin{split}
\b_2 &= -\lambda^3 \frac{11}{11520} \chi\bs{-2}^3\\
\b_3 &= - \lambda^6 \frac{2981}{1741824000}\chi\bs{-2}^5\,.
\end{split}
\ee
Notice that these beta functions are again specific to the toy $\Phi^3$ model and furthermore they are not determined by conformal invariance alone. Therefore, their coefficients cannot be compared with computations in a dual conformal field theory.

\section{Conclusions and generalization}
\label{sec:concl}
In the previous section we have obtained the holographic Callan-Symanzik equation for a simple scalar $\lambda \Phi^3$ toy model. We however believe that the structures we exhibited are more general. In particular, we may directly generalize our method of deriving the holographic Callan-Symanzik equation to the situation with multiple (not necessarily scalar) bulk fields as follows.

Consider a regularized bare on-shell bulk action which depends on the boundary values of a generic set of bulk fields $\Phi^I$. (This set of bulk fields implicitly includes the metric $\g_{ij}$ on which the on-shell action always depends.) The counterterm action splits into the single-trace counterterm action $S_\text{ct}^\text{st}$, which define renormalized conjugate momenta $\Pi_r^I$ via
\be
\delta (S_\text{on-shell} + S_\text{ct}^\text{st}) = \int d^d x \sqrt{\g} \sum_I \Pi_r^I \delta \Phi^I
\ee
and the multi-trace counterterm action $S_\text{ct}^\text{mt}$ which contains terms which are at least quadratic in the $\Pi_r^I$. After adding all the counterterms the action remains by construction finite as $r \to \infty$. 

The first variation of the on-shell action, given in principle by:
\be
\delta(S_\text{on-shell} + S_\text{ct}^\text{st} + S_\text{ct}^\text{mt}) = \int d^d x \sqrt{\g} (\ldots)\delta \Phi^I + \int d^d x \sqrt{\g} (\ldots) \delta \Pi_r^I \,,
\ee
takes as $r \to \infty$ the form (repeated indices are summed over):
\be
\begin{split}
\label{eq:deltaSren}
\delta S_{\text{ren}} &= \lim_{r \to \infty} \delta (S_{\text{on-shell}} + S_\text{ct}^\text{st} + S_\text{ct}^\text{mt}) =
\int d^d x \sqrt{g} (\pi^I \delta \phi^I\bs{\Dm} + \tilde \pi^I \delta \phi^I\bs{\Dp}) \\
&= \int d^d x \sqrt{g} \pi^I_\chi \delta \chi^I\bs{\Dm}\,,
\end{split}
\ee
where the $\phi^I\bs{\Dpm}$ are the usual coefficients of the radial expansion of the field $\Phi^I$ at order $\exp(-\Dpm r)$ which at the free-field level correspond to the source and the vev term, respectively. If $\tilde \pi^I$ were zero, then the one-point functions would be directly defined by the $\pi^I$. On the other hand, if $\tilde \pi^I$ does not vanish there is mixing between single- and multi-trace operators. This occurs only if some of the multi-trace counterterms are logarithmic in nature. In such cases we perturbatively redefine the source as in \eqref{eq:tildephiab0dm} such that the variation is of the form displayed on the last line of \eqref{eq:deltaSren} for some new source $\chi^I\bs{\Dm}$. (Alternatively we may choose a convenient renormalization scheme as we did in subsection \ref{sec:thirdorder} but in general we prefer not to fix the scheme dependence.) In that case the one-point functions of the renormalized operators $\op^I$ are given by the $\pi^I_\chi$.

The holographic Callan-Symanzik equation is obtained as:
\be
\lim_{r \to \infty} \frac{d}{dr} \Big( S_{\text{on-shell}} + S_{\text{ct}}^\text{st} + S_\text{ct}^\text{mt} \Big) = 0\,.
\ee
Using the method explained in the previous section we may rewrite this as:
\be
\label{eq:genCS1}
0 = \lim_{r \to \infty} \Big(\delta_r +  \frac{\del}{\del r}\Big) (S_{\text{on-shell}} + S_{\text{ct}}^\text{st} + S_\text{ct}^\text{mt})\,,
\ee
with $\delta_r$ defined as
\be
\delta_r \equiv \int \dot \Phi^I \fdel{\Phi^I} + \int \dot \Pi_r^I \fdel{\Pi^I}\,.
\ee
We then evoke the reasoning of subsection \ref{sec:higherorderst} to write \eqref{eq:genCS1} as:
\be
\label{eq:generalCShol}
\begin{split}
0 &= \lim_{r \to \infty}  \int \Big(\delta_r \phi^K\bs{\Dm}\fdel[S_\text{ren}]{\phi^K\bs{\Dm}} + \delta_r \phi^K\bs{\Dp} \fdel[S_\text{ren}]{\phi^K\bs{\Dp}} \Big) + \lim_{r \to \infty} \frac{\del}{\del r} (S_{\text{ct}}^\text{st} + S_\text{ct}^\text{mt})\\
&= \int \Big((- \Dm \phi^K\bs{\Dm}+ \tilde \phi^K\bs{\Dm})\pi^K +(- \Dp \phi^K\bs{\Dm}+ \tilde \phi^K\bs{\Dp}) \tilde \pi^K    \Big) + \lim_{r \to \infty}\frac{\del}{\del r}  (S_{\text{ct}}^\text{st} + S_\text{ct}^\text{mt})\,,
\end{split}
\ee
where $\tilde \phi^I\bs{\Dpm}$ are the coefficients of $r \exp(- \Dpm r)$ in the radial expansion of the field $\Phi^I$, respectively. Notice that all terms in \eqref{eq:generalCShol} are explicit functions of $\phi^I\bs{\Dm}$ and $\phi^I\bs{\Dp}$. The final form of the general holographic Callan-Symanzik equation is now obtained by changing variables to $\chi^I\bs{\Dm}$ and $\pi^I_\chi$. The terms which are of quadratic or higher order in $\pi^I_\chi$ are then interpreted in terms of multi-trace operator mixing following the reasoning of subsection \ref{sec:multitrace}.

The arguments presented above demonstrate that equation \eqref{eq:generalCShol}, interpreted as a function of $\chi^I\bs{\Dm}$ and $\pi^I_\chi$, is the most general form of a Callan-Symanzik equation that we may obtain holographically. We therefore see it as the main result of our paper.

Let us emphasize that \eqref{eq:generalCShol} has precisely the expected form for a most general Callan-Symanzik equation of a large $N$ field theory as follows from equation \eqref{eq:generalcseqn} combined with \eqref{eq:mddmW} and the multi-trace results of section \ref{sec:multitrace}. Another non-trivial check of this result is the fact that the asymptotic analysis in the bulk suffices to obtain the complete Callan-Symanzik equation, including all the beta functions. This matches the results obtained in the holographic renormalization literature and reflects the usual separation of scales in quantum field theory.

The results we presented in this paper can be extended in various directions. First of all, it would be interesting to obtain the conformal Ward identity corresponding to general local Weyl rescalings in the presence of multi-trace counterterms. This can be done by inclusion of gravity in the bulk. One concrete example would be to extend the model outlined in section 3 of \cite{me} to a case which involves logarithmic divergences. It would also be very interesting to extend the general Hamilton-Jacobi approach to holographic renormalization \cite{deBoer:1999xf,deBoer:2000cz,Martelli:2002sp,Papadimitriou:2004ap} to include sources for irrelevant operators. This may shed some light on the mechanism which underlies the absence of counterterms linear in the conjugate momentum. Finally, our results will likely be useful in the study of non-AlAdS spacetimes, not only for renormalizing dual field theory correlation functions but also for the related investigation of asymptotic symmetries and the corresponding  definition of conserved charges.

\section*{Acknowledgments}
We would like to thank Leonardo Rastelli and Kostas Skenderis for their comments on the draft.

\appendix
\section{Multi-trace operators at large $N$}
\label{sec:multitracelargeN}
In this section we will review the standard prescription for multi-trace deformations at large $N$, following \cite{Berkooz:2002ug,Witten:2001ua,Mueck:2002gm,Sever:2002fk,Elitzur:2005kz,Papadimitriou:2007sj}. We suppose that our quantum field theory obeys the same large $N$ counting rules as for example Yang-Mills theory in the 't Hooft limit. In terms of the fundamental fields $\Phi^i$ of the theory, our composite single-trace operators are defined as:
\be
\op^I = \Tr(\Phi^{i_1} \Phi^{i_2} \ldots \Phi^{i_k})\,,
\ee
where the abstract index $I$ is used to label the different operators in the theory. If we keep the 't Hooft coupling fixed then the action $S$ takes the schematic form:
\be
S = N \sum_I c_I \op^I\,,
\ee
where the coefficients $c_I$ are $O(N^0)$. This action results in a factor of $N^{-1}$ for each propagator and a factor of $N$ for each vertex and loop. Following 't Hoofts arguments we therefore find that the single-trace partition function, defined as
\be
\exp(- W_0[t^I]) = \int D\Phi^i \exp\Big(-S - N \int t	I \op_I \Big) \,,
\ee
scales to leading order as $N^2$. Let us rewrite it as:
\be
 W_0[t^I] = N^2 w_0[t^I]
 \ee   
so $w_0[t^I]$ is of order $N^0$. Upon functional differentiation with respect to the $t^I$ we also observe that $n$-point correlation functions of the operators $\op_I$ scale as $N^{2-n}$.\footnote{One may of course introduce rescaled operators $\hat \op(x) = N \op(x)$ whose $n$-point correlation functions all scale as $N^2$. This would also be the normalization one would obtain from a canonically normalized on-shell supergravity action. We however choose not to implement this rescaling here.}

Let us now consider switching on a multi-trace interaction. More precisely, we choose our factors of $N$ such that the multi-trace partition function takes the form:\footnote{One may of course introduce rescaled operators $\tilde \op(x) = N^{-1} \op(x)$ in order to absorb the explicit $N^{-1}$ in equation \eqref{eq:multitracepartfn}. This would also be the normalization considered in much of the multi-trace literature, see for example \cite{Witten:2001ua}. We however choose not to implement this rescaling here.}
\be
\label{eq:multitracepartfn}
\exp(- W[t, f]) = \int D\Phi \exp\Big(- S - N \int t \,\op - N^2 \int f(N^{-1} \op) \Big) \,,
\ee
where $f(\s)$ is a polynomial in $\s$ with order one coefficients and for simplicity of notation we removed the indices $I$ and $i$ labelling the operators and the fundamental fields. We now formally rewrite this partition function as:
\be
\begin{split}
&\exp(- W[t, f]) = \int D\Phi \exp\Big(- S - N \int t\, \op - N^2 \int f(N^{-1} \op) \Big)\\
&\quad= \int D\Phi D\a D\s \exp\Big(-S - N \int t\, \op - N^2 \int f(\s)  - N^2 \int \a ( \s - N^{-1} \op) \Big)\\
&\quad= \int D\a D\s \exp\Big(- N^2 w_0[t - \a] - N^2 \int (f(\s) + \a \s) \Big)\\
&\quad\sim \exp\Big( - N^2 w_0[t + f'(\s)] - N^2 \int (f(\s) - f'(\s)\s) \Big)\,,
\end{split}
\ee
where the saddle-point approximation in the last line also dictates that:
\be
\label{eq:smultitrace}
\s = w'_0[t + f'(\s)]\,.
\ee
We conclude that, to leading order in the large $N$ expansion:
\be
\label{eq:wmultitrace}
w[t,f] = w_0[t + f'(\s)] + \int d^d x (f(\s) - f'(\s) \s)\,,
\ee
with $\s$ as above and all terms in sight are now of order one.

\subsection{Correlation functions involving multi-trace operators}
At large $N$ the correlation functions involving multi-trace operators are easily seen to factorize by inspection of the relevant Feynman diagrams. This leads to for example:
\be
\vev{\op^2(y) \op(x_1) \op(x_2) \op(x_3)} = 2 \vev{\op(y)\op(x_1)} \vev{\op(y)\op(x_2)\op(x_3)} + (\text{2 permutations})\,.
\ee
Let us now investigate to which extend the partition function $W[t,f]$ is a generating functional of these factorized correlation functions. To this end we write
\be
f(\s(x)) = \sum_{k \geq 2} t_k(x) \s(x)^k\,,
\ee
so from \eqref{eq:multitracepartfn} we see that the $t_k(x)$ play the role of sources for the multi-trace operators $\op^k(x)$. We will also use the notation $t_1(x) = t(x)$ for the source of the single-trace operator $\op(x)$. Let us now consider a small fluctuation of all the sources:
\be
t_k(x) \to t_k(x) + \delta t_k(x)\,.
\ee
Using the above definitions we find that all the variations involving $\delta \s$ cancel and we eventually obtain:
\be
\label{eq:deltawmultitrace}
\delta w[t_k] = \sum_{k \geq 1}\int d^d x \, \s(x)^k \delta t_k(x) \,.
\ee
According to \eqref{eq:multitracepartfn} we find that functional differentiation with respect to $t$ generates insertions of $N \op$ whereas functional differentiation with respect to the other $t_k$ generates insertions of $N^{2-k} \op^k$. Reinstating those factors of $N$ as well as the overall $N^2$ of the partition function, we obtain that
\be
\vev{\op^k(x)} = N^k \s(x)^k \qquad \qquad \text{for all }k \geq 1\,.
\ee
Notice that upon further functional differentiation we find for example that:
\be
\label{eq:multitracecorr}
\begin{split}
\vev{\op^2(x) \op^2 (y)} &= N^2 \Big(\fdel{t_2(x)} \fdel{t_2(y)} w[t_k]\Big)\Big|_{t_k = 0}  = 0\\
\vev{\op^2(x) \op(y) \op(z)} &= \Big(\fdel{t(z)} \fdel{t(y)} \fdel{t_2(x)} w[t_k]\Big)\Big|_{t_k = 0} = 2 \vev{\op(x)\op(y)} \vev{\op(x)\op(z)}\,,
\end{split}
\ee
where we used that $\vev{\op(x)}|_{t_k = 0} = 0$ by conformal invariance. Perhaps surprisingly, the deformed partition function does not capture the double-trace two-point function but it does capture the three-point function involving one double-trace and two single-trace operators. The reason for this lies in the extra $N^2$ on the first line of \eqref{eq:multitracecorr}. To see this, notice that to leading order we know that both of the above correlators should factorize into a product of two single-trace two-point functions and they are therefore necessarily both of order $N^0$ (recall that single-trace $n$-point functions scale as $N^{2-n}$). However, the extra $N^2$ on the first line in \eqref{eq:multitracecorr} implies that the partition function gives us the order $N^2$ term in the two-point function of $\op^2$. This should evaluate to zero and indeed we find that it does. On the other hand, there is no such factor on the second line of \eqref{eq:multitracecorr}. We are thus computing the $N^0$ term in the three-point function and indeed we obtain a non-zero answer.

The above result can be generalized as follows. As we mentioned above, functional differentiation with respect to the sources $t_k$ actually generates insertion of $N^{2-k} \op^k$. On the other hand, the only $N$-dependence in the partition function is an overall $N^2$. Therefore the deformed multi-trace partition function captures only those correlation functions for which:
\be
\vev{N^{2-k_1} \op^{k_1} N^{2-k_2} \op^{k_2} \ldots N^{2 - k_n} \op^{k_n}} \sim N^2 .
\ee
In other words, only if a correlation function satisfies:
\be
\label{eq:multitraceinpartnfn}
\vev{\op^{k_1} \op^{k_2} \ldots \op^{k_n}} \sim N^{2 - 2n + \sum_{i=1}^n k_i}
\ee
is it captured by the partition function. Upon substitution of the correct values of the $k_i$ and $n$ we again recover the results in \eqref{eq:multitracecorr}.

\bibliographystyle{utphys}
\bibliography{biblio}

\end{document}